 \documentclass[preprint2]{aastex}

\usepackage{amsmath}
\usepackage{amssymb}
\usepackage{amsfonts}
\newcommand{\udeg}[0]{\ensuremath{^{\circ}}}                 
\newcommand{\fmin}[0]{\hbox{\ensuremath{.{\mkern -4mu}'}}}                       
\def\sun{\odot}
\def\url{\texttt}

\DeclareRobustCommand{\ion}[2]{%
\relax\ifmmode
\ifx\testbx\f@series
{\mathbf{#1\,\mathsc{#2}}}\else
{\mathrm{#1\,\mathsc{#2}}}\fi
\else\textup{#1\,{\mdseries\textsc{#2}}}%
\fi}
\DeclareRobustCommand{\tempion}[2]{%
\relax\ifmmode
\ifx\testbx\f@series
{\mathbf{#1\mathsc{#2}}}\else
{\mathrm{#1\mathsc{#2}}}\fi
\else\textup{#1{\mdseries\textsc{#2}}}%
\fi}

\shorttitle{EBHIS: Data reduction}
\shortauthors{Winkel et al.}

\begin{document}


\title{The Effelsberg--Bonn \ion{H}{i} Survey: Data reduction}


\author{B. Winkel, P.\,M.\,W. Kalberla, J. Kerp, and L. Fl\"{o}er}
\affil{Argelander-Institute for Astronomy (AIfA), University of Bonn,\\  Auf dem H\"{u}gel 71, D-53121 Bonn, Germany}
\email{bwinkel@astro.uni-bonn.de}




\begin{abstract}
Starting in winter 2008/2009 an \textit{L}-band 7-Feed-Array receiver is used for a 21-cm line survey performed with the 100-m telescope, the Effelsberg--Bonn \ion{H}{i} survey (EBHIS). The EBHIS will cover the whole northern hemisphere for decl.$ > -5\degr$ comprising both the galactic and extragalactic sky out to a distance of about 230\,Mpc. Using state-of-the-art FPGA-based digital fast Fourier transform spectrometers, superior in dynamic range and temporal resolution to conventional correlators, allows us to apply sophisticated radio frequency interference (RFI) mitigation schemes.

In this paper, the EBHIS data reduction package and first results are presented. The reduction software consists of RFI detection schemes, flux and gain-curve calibration, stray-radiation removal, baseline fitting, and finally the gridding to produce data cubes.
The whole software chain is successfully tested using multi-feed data toward many smaller test fields (1--100 deg$^2$) and recently applied for the first time to data of two large sky areas, each covering about 2000 deg$^2$. The first large area is toward the northern galactic pole and the second one toward the northern tip of the Magellanic Leading Arm. Here, we demonstrate the data quality of EBHIS Milky Way data and give a first impression on the first data release in 2011.
\end{abstract}





\section{Introduction}
Blind \ion{H}{i} surveys allow us to explore a variety of objects. The Effelsberg--Bonn \ion{H}{i} Survey (EBHIS) covers not only the Milky Way but also the local universe out to a redshift of 0.07.
Accordingly, EBHIS will serve as a major database for many scientific questions, addressing, e.g., the structure and mass distribution of the Milky Way, the size distribution of halo clouds, or the structure formation of the local universe. A complete description of the scientific aims will be given in a forthcoming paper. Here, the technical setup and the data reduction software as used for the survey are presented.

Today, the most comprehensive database for galactic \ion{H}{i} science is the Leiden/Argentine/Bonn survey \citep[LAB;][]{kalberla05} the first all-sky survey corrected for stray radiation. Toward high galactic latitudes, the faint Milky Way emission from the direction of interest can be severely degraded by radiation of the Milky Way disk entering the receiver system via the near  and far side lobes. SR correction is crucial for quantitative analyses of most of the galactic sky using conventional single dish optics.  
The Parkes and Arecibo telescopes perform multiple consecutive surveys to observe the galactic and extragalactic sky. The data acquisition of the Parkes Galactic All Sky Survey \citep[GASS;][]{mcclure06,mcclure09} is already completed and the final data products are released \citep{kalberla10}. The Galactic ALFA \citep[GALFA;][]{goldsmith04,heiles04} is still in progress. In the extragalactic regime the \ion{H}{i} Parkes All Sky Survey \citep[HIPASS;][]{barnes01,meyer04} mapped the complete southern hemisphere detecting more than 5000 galaxies providing a valuable database to infer the \ion{H}{i} properties of galaxies in the local universe. The Arecibo Legacy Fast ALFA survey \citep[ALFALFA;][]{giovanelli05p} is ongoing and will result in deeper data at higher angular resolution though it is limited to a smaller area of about 7\,000 deg$^2$.

Since the beginning of the year 2009 an \textit{L}-band 7-Feed-Array receiver system is available at the 100-m telescope in Effelsberg for astronomical measurements. We use this instrument to perform a fully sampled \ion{H}{i} survey of the northern hemisphere for $\mathrm{decl.}> -5\degr$, observing in parallel the galactic and extragalactic sky. Multiple coverages (at different hour angles and seasons) will allow us to disentangle radio frequency interference (RFI), SR, and baseline problems. Redundancy allows us to remove instrumental biases significantly. The survey area is subdivided into two sky areas, one toward the Sloan Digital Sky Survey \citep[SDSS;][]{adelman08} area where an integration time of 10\,minutes is projected while the remaining sky will be integrated for 2\,minutes. This yields a full-sky survey superior in sensitivity, angular sampling, and resolution to any previously performed large \ion{H}{i} survey \citep{kalberla09}. The data of the first full sky coverage will be released in 2011.

The data analysis of a large-area \ion{H}{i} survey is undoubtedly a major challenge. Modern receiving systems offer scientifically usable bandwidths of several tens to hundreds of MHz to study the \ion{H}{i} at moderate red shifts. Digital high-dynamic-range spectrometers based on Field Programmable Gate Arrays \citep[FPGAs; ][]{stanko05,klein06} allow us to store spectra on time scales of less than 1\,s which is essential to apply sophisticated RFI mitigation procedures \citep{winkel07a}. The receiving system as used for EBHIS provides a bandwidth of 100\,MHz which covers the redshift range out to $z\sim0.07$ while the 16384 spectral channels yield a velocity resolution of $1.25\,\mathrm{km\,s}^{-1}$. Spectral dumps are recorded every 500\,ms.
As a drawback though, one has to deal with very high data rates. For the EBHIS, typical values are about 5\,GB\,hr$^{-1}$, making the calibration and analysis of EBHIS data a task which is impossible to accomplish manually. Consequently, the data reduction algorithms and software developed were optimized with respect to computational efficiency and performance.

The paper is organized as follows. In Section\,\ref{secinstrument}, the receiving system and technical setup are described. During various test measurements the quality of the new receiving system was investigated, the results of which are discussed in Section\,\ref{secreceiverquality}. Our data reduction package which was developed for the EBHIS is presented in detail in Section\,\ref{secdatareduction} and we show its successful application to an example data set in  Section\,\ref{secfirstresults}. A summary is given in  Section\,\ref{secsummary}.

\section{Technical setup}\label{secinstrument}
\subsection{Receiver}\label{subsecreceiverandpattern}

\begin{figure}[!t]
\centering
\includegraphics[width=0.45\textwidth,clip=,bb=55 50 405 291]{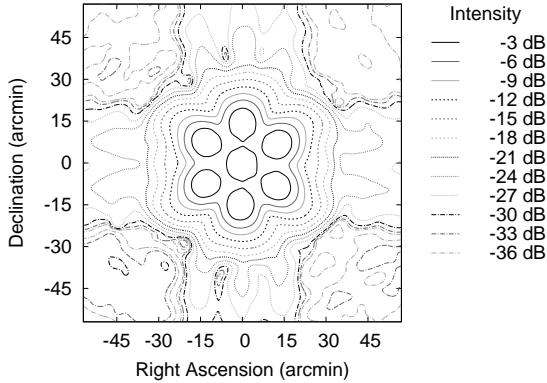}
\caption{Antenna pattern of the seven-beam receiver as measured by E. F\"{u}rst (MPIfR) using the continuum backend. The contour lines mark the sensitivity distribution (in dB) of the receiver system as a function of angular distance from the source. The cross-like structure is caused by the feed-support legs of the primary focus cabin. Obviously, the off-set feeds are affected by coma, leading to the obvious elongation of the sensitivity pattern.}
\label{figsevenbeampattern}
\end{figure}

The 21-cm multi-feed receiver is a cooled single conversion heterodyne system offering 14 separate receiving channels. The electromagnetic waves are focused by the antenna and coupled via the feed horns into waveguides placed in a cryogenic dewar which is situated in the primary focus of the telescope. The central feed is sensitive for circular polarization while the off-set feeds receive only linear polarized signals. According to this setup, the offset feeds are expected to be more sensitive for a broader variety of RFI signals.
An initial amplification of 40\,dB is applied to the spectral band between 1200\, and 1700\,MHz. A lot of effort went into arranging the beams providing best possible beam efficiency while minimizing inter-beam coupling. Before down-conversion to the intermediate frequency (IF) using a local oscillator (LO) further filtering is applied, limiting the bandwidth to the range $1290-1430\,\textrm{MHz}$. The IF band lies between 80\,MHz and 220\,MHz (the center frequency is 150\,MHz). The receiver utilizes a noise diode at constant temperature to monitor relative modulations of the system temperature. 
The beam separation is $15\arcmin$, or 1.6 beam widths, placed on a hexagonal grid. In Figure\,\ref{figsevenbeampattern}, the measured beam pattern for the multi-feed array is shown.  A more detailed description can be found in \citet{keller06techreport}.

\subsection{Backend}\label{subsecbackend}
For the EBHIS a fast Fourier transform (FFT) spectrometer developed at the Max-Planck-Institut f\"{u}r Radioastronomie (MPIfR) is used. It is based on a Xilinx Virtex-4 FPGA, fed by an analog--digital converter (ADC) that has a sampling rate up to 3\,GHz at 8 bit dynamics. The original design was introduced for molecular spectroscopy at the APEX telescope \citep{klein06}. It was adapted to match the desired specifications for an \ion{H}{i} survey. Since 2008 August seven backends, each equipped with two spectral channels, are installed.
They utilize a total bandwidth of 100\,MHz and 16k spectral channels, providing an effective frequency resolution of 7.1\,kHz \citep[equivalent noise bandwidth, ENBW;][]{klein08}, yielding a velocity resolution of 1.45\,km\,s$^{-1}$, which is only marginally larger than the channel separation of 6.1\,kHz. Every 500\,ms an integrated spectrum (abbreviated henceforward as ``spectral dump'') is transmitted via Ethernet to a server, which stores the raw data (in binary file format) in the disk.  

\subsection{Survey strategy}
The survey will be carried out in three major steps. The first part, the testing phase, is already finished. Here, we mapped smaller portions of the sky to test the instrument and software. Currently, the complete northern hemisphere is observed with an effective integration time of 2\,minutes per position. Finally, the SDSS area will be observed via multiple coverages yielding the finally aimed integration time of 10\,minutes.
$5\times5$ degree fields are measured with on-the-fly R.A.--Decl. scanning using a tangential plane projection which will lead to a homogeneous noise distribution all over the sky. The hexagonal feed pattern is rotated by $19\deg$ with respect to the scanning direction such that the sampled scanlines (denoted as subscans) are equidistantly spaced. To keep this feed angle fixed in the R.A.--Decl. system, the dewar must be rotated according to the parallactic angle.

\section{Receiver quality}\label{secreceiverquality}
Several test measurements have been conducted, not only to test the new multi-beam receiver and the FFT spectrometers but also to check the quality of the data reduction software. The first test measurements (2007 November 20/21 and 23/24) were intended to investigate the overall system performance by carrying out \textit{Allan} tests, measurements of the system temperature, and bandpass stability. During these observations a strong RFI signal produced by terrestrial digital radio (DAB) at a frequency of 1450\,MHz was observed (located in frequency outside of the bandpass filter). This strong terrestrial irradiation caused highly variable baselines for the central (circularly polarized) feed. As a consequence, new stop-band filters were installed, having much higher sideband suppression. After that we identified a significant feed resonance producing a strong broad Gaussian-like signal occurring at a frequency of about 1435\,MHz. To minimize its influence the LO setup was optimized accordingly. The emission of the Milky Way is detected close to the high-frequency limit of the bandpass edge. 

\begin{figure}[!t]
\centering
\includegraphics[width=0.45\textwidth,clip=,bb=66 45 410 270]{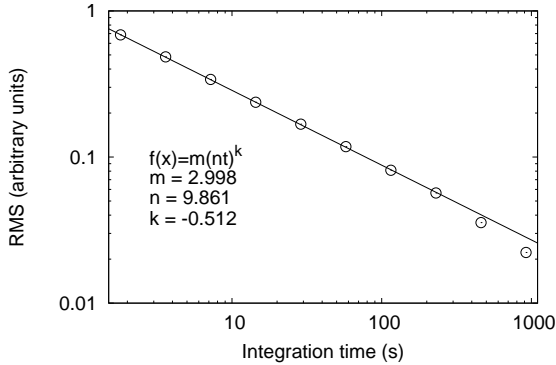}
\caption{Allan plot calculated for spectral data, as observed  during test measurements in 2007 November. The  plot shows the noise behavior of one of the offset feeds. Each spectral dump is integrated for 450\,ms (separated in time by 2\,s). The plot reveals very good receiver stability up to the total integration time of about 1000\,s.}
\label{figtm1allanplot}
\end{figure}

It is very important to show that the noise is free from systematic effects. For an ideal radio receiver the radiometer equation is applicable
\begin{equation}
P(t)\sim \frac{1}{\sqrt{t\cdot\Delta f}}\label{eqradiometer2}
\end{equation}
stating that the noise power $P(t)$ decreases with the square root of the integration time $t$ if a constant bandwidth $\Delta f$ is used. Practically, each system suffers from instabilities on a certain timescale $t_\mathrm{A}$, yielding a divergence from the theoretical behavior. It is obvious that a system should be designed in a way to maximize $t_\mathrm{A}$. One possibility of determining $t_\mathrm{A}$ for the whole receiving system is to compute an Allan plot. Several thousands of spectral dumps were recorded and subsequently integrated (e.g., in steps of $2^n$) to evaluate the noise behavior on the timescale $t_n$. 
In Figure\,\ref{figtm1allanplot}, an Allan plot for one of the offset feeds is shown. The data show the expected behavior up to an integration time of about 1000\,s, which is sufficient for the deepest observations (600\,s) we are aiming for.

\begin{figure}[!t]
\centering
\includegraphics[scale=0.6,clip=,bb=66 80 407 291]{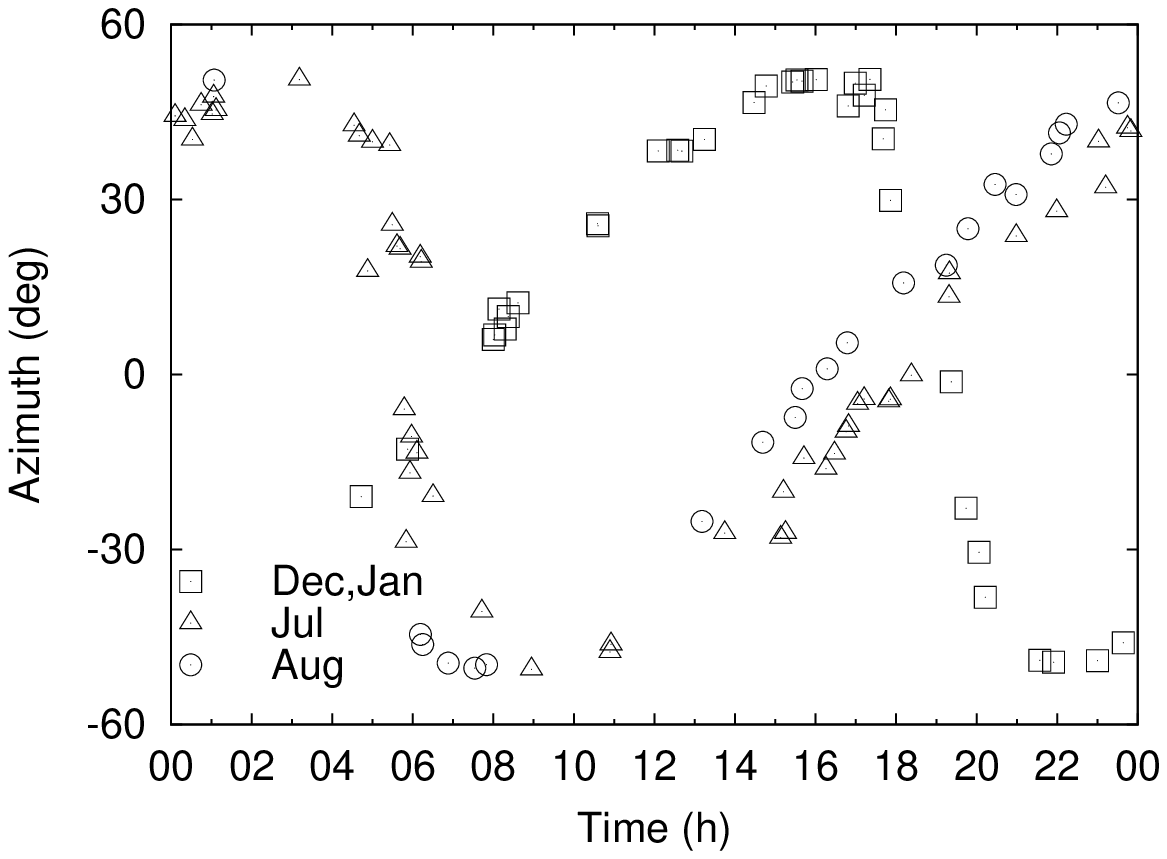}\\[1ex]
\includegraphics[scale=0.6,clip=,bb=66 80 407 291]{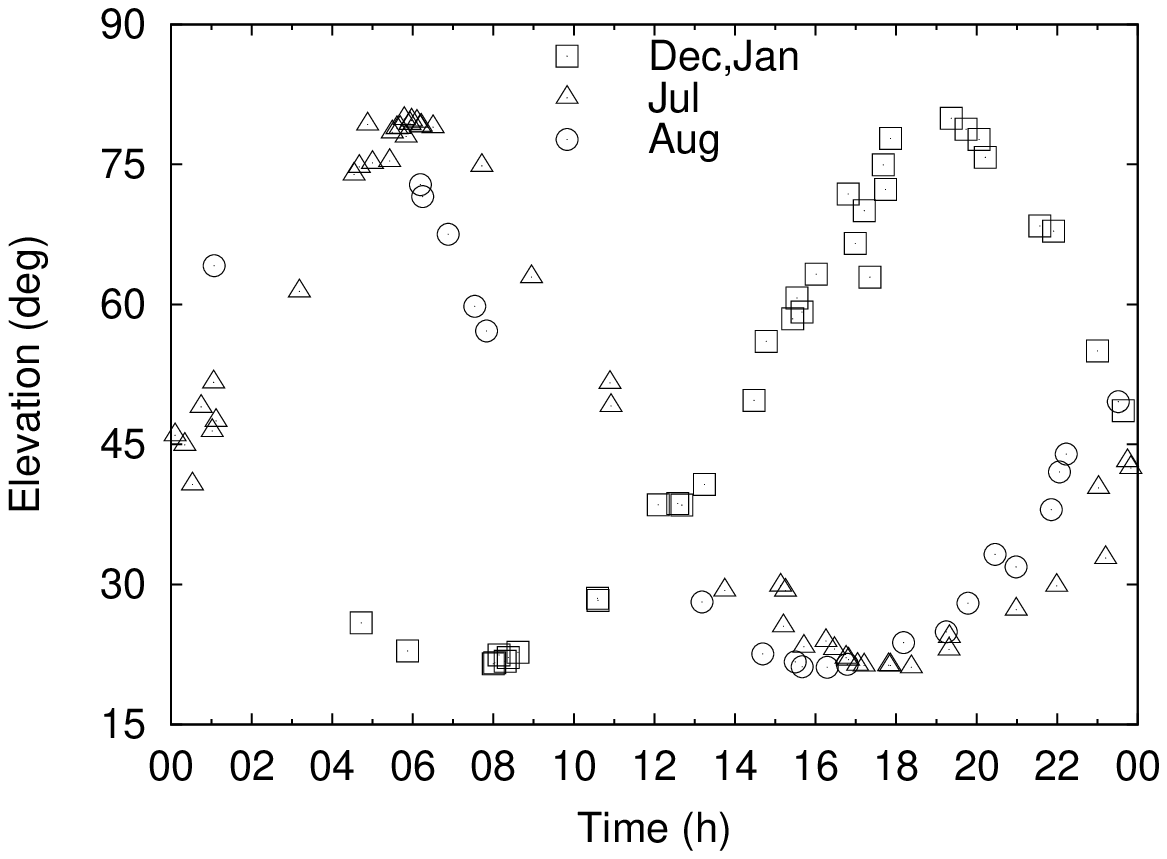}\\[1ex]
\includegraphics[scale=0.6,clip=,bb=66 50 407 291]{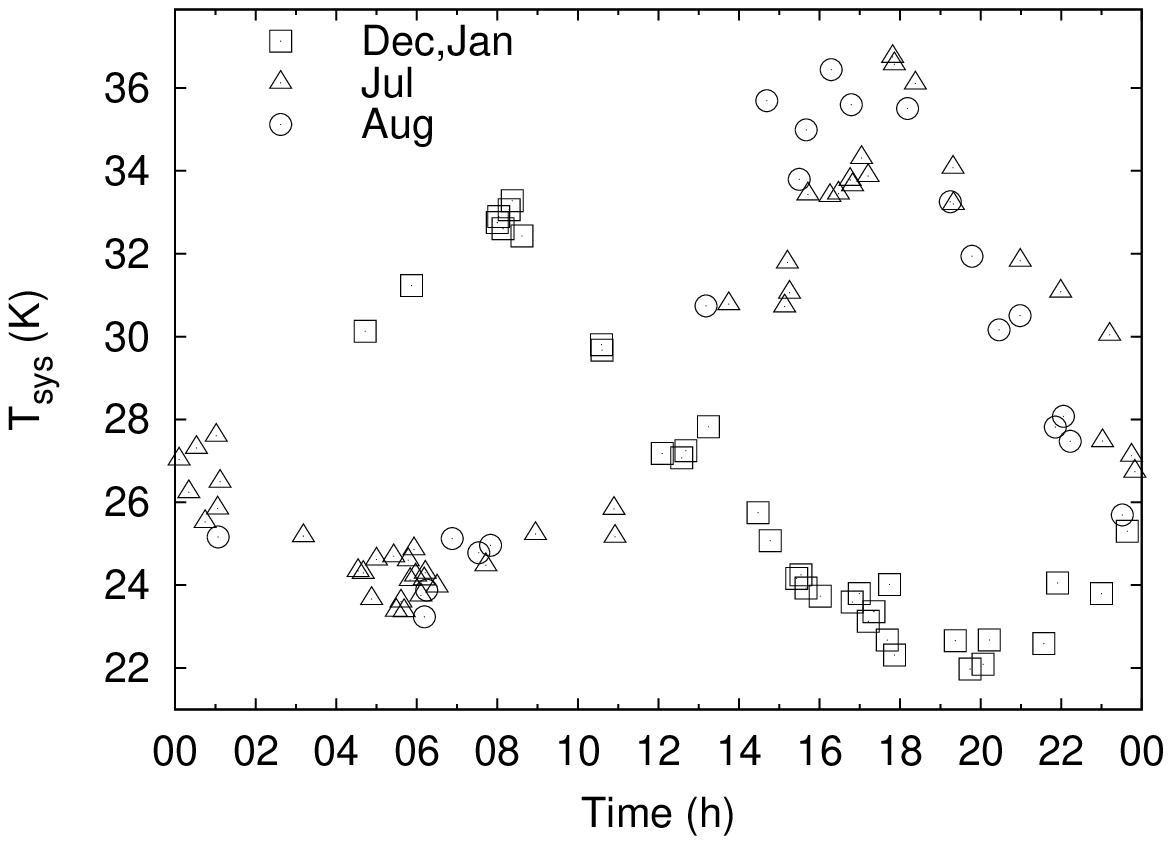}
\caption{System temperatures obtained using the IAU standard calibration source S\,7 for three different observational periods. In the upper and middle panels, the horizontal coordinates of S\,7 are plotted as a function of time. The lower plot contains the resulting system temperatures. }
\label{figtsys}
\end{figure}

In Figure\,\ref{figtsys} (lower panel), the system temperatures for the central feed are shown as derived from observations of the standard calibration source S\,7. Data for three different observational periods are plotted separately. Figure\,\ref{figtsys} shows two main effects; as expected $T_\mathrm{sys}$ is strongly elevation dependent, i.e., lower elevations lead to higher irradiation from the ground (Figure\,\ref{figtsys}, middle panel), and seasonal effects are apparent. The maximal system temperature is higher during summer than during the winter term.  The remaining scatter can be attributed to local weather conditions and the averaging of the data over a longer period (about a month). The system temperature of the central feed ranges typically between 22 and 35\,K. The corresponding temperatures of the offset feeds are marginally higher, ranging between 24\,K and 39\,K. Note, that the apparently high system temperatures in some of the measurements can be attributed entirely to the low elevation of the calibration source.

\section{The EBHIS data reduction}\label{secdatareduction}

The major reduction tasks are flux calibration, gain curve (bandpass) correction, SR correction, baseline fitting, and RFI detection. The processed spectra are eventually merged by a gridding tool into a three-dimensional data cube.
Deviating from the common pipeline approaches, the EBHIS reduction is organized in a highly flexible manner. To minimize redundancies, each  processing step works as independently as possible on the data. To achieve this goal, every correction to be applied on the data is described by a minimal set of  parameters which is then stored in an SQL database. For example, the RFI detection algorithm returns a list of spectral channels per  spectral dump. Only this short list consisting of a few integer entries needs to be stored. Every other processing software can then easily query the database to obtain these lists and flag contaminated data. Also the gain calibration needs only few numbers to store per observation session. Tasks, which rely on calibrated input spectra, simply read the raw data from disk and apply the calibration on-the-fly. 
\begin{figure}[!t]
\centering
\includegraphics[width=0.45\textwidth]{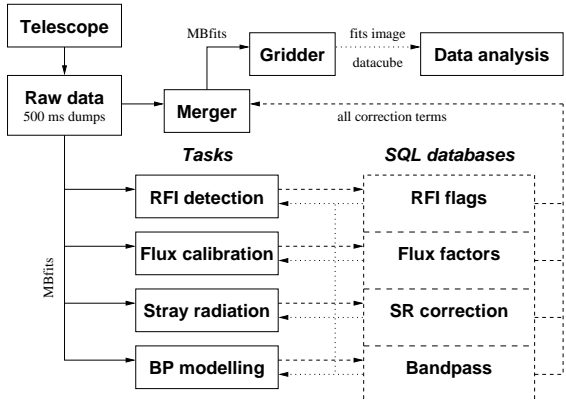}
\caption{Data reduction scheme used for EBHIS is organized in a highly flexible way. Each task is optimized to work as independently as possible on the individual spectral dumps. The computed correction parameters are stored within an SQL database to allow an easy query of necessary parameters (e.g., calibration factors, RFI flags) from other tasks.}
\label{figdatareductionscheme}
\end{figure}

If the individual modules do not depend on each other, this modular approach of the data reduction chain enables to update one single module without the  necessity to re-execute other reduction tasks. In that case, different tasks can even work in parallel.
A schematic representation of the EBHIS data processing chain is shown in Figure\,\ref{figdatareductionscheme}. The picture shows with solid line arrows the flow of the spectral data, while dashed lines show the information exchange with the databases. Every task stores calculated values in an associated database table, and can query all other databases as necessary. 
In contrast to pipeline approaches, a new task is necessary, which is denoted as ``merger'', to apply all the correction terms to generate the final data set.  After that, the spectra can be gridded to a data cube.

As the integration time per dump is $500\,\mathrm{ms}$, a significant amount of data has to be handled for the full survey (about 25\,Tbyte).
We developed serial algorithms, to allow processing of large data sets without the need to keep them in the main memory (RAM), which is especially important for the gridding task.
Multi-threading techniques are extensively used to significantly speed up computation on multi-processor/-core platforms. Further improvement on processing speed is because of the database approach which allows the simultaneous access from multiple workstations. The database also provides the opportunity to easily query information, e.g., on sky areas already observed (including visualization), and allows to backup the database very efficiently.

A typical work flow is as follows. RFI detection, flux, and gain curve calibration are independent, they can be processed in parallel. From the calibrated spectra, the SR correction is subtracted, baselines are computed, and finally the gridder computes the data cubes. However, an accurate SR correction needs in principle an iterative approach, where the resulting data of one iteration are fed into an improved SR model to be subtracted in the next iteration.

\subsection{RFI mitigation}\label{subsecrfidetection}

In radio astronomy, the signals of interest are in general polluted by man-made artificial radio emission called RFI. RFI is produced by a broad variety of emitters. Because of the high sensitivity of modern radio telescopes, faintest RFI irradiation can corrupt the observational data. Even within the radio astronomical protected bands one has to deal with ``legal'' RFI irradiation in adjacent spectral regimes, leaking part of their radiation power into the protected band below the irradiation threshold.
Digital circuits, common in modern electronic devices, pollute their environment with a multitude of harmonics. The dynamic range of RFI irradiation covers a huge dynamic range from events close to the statistical noise level up to strong bursts which can even lead to the saturation of the whole receiving system.

Two practical strategies are used today to deal with RFI: keeping artificial radiation away from the telescope (passive mitigation) or trying to detect or even mitigate RFI once it has entered the receiving system (active mitigation). For the latter, several methods have been developed so far.  First, a manual search for RFI signals in the spectra can be carried out (the ``classical'' method).  Second, using sophisticated algorithms it is possible to search for RFI signals in recorded data automatically \citep{bhat05}. Third, one can use real-time applications which have to be implemented into the receiver chain of the telescope. Various approaches are under consideration, e.g., adaptive filters \citep{bradleybarnbaum96}, post-correlators \citep{briggs00}, or real-time higher order statistics \citep[HOS; see][]{fridman01}.

For EBHIS we follow the second approach as no hardware solution is available at the 100-m telescope.
It became obvious during test observations \citep{winkel07a} that the RFI amplitude variations at Effelsberg occur either on the order of less than $100\,\mathrm{ms}$ or the amplitudes are relatively stable. To increase the detection rate for the former, spectral data at high temporal resolution are desired. The necessary short integration times yield very high data rates and low signal-to-noise spectra. In practice, one has to find a compromise between reasonable time resolution for RFI detection and the amount of recorded data. Today, only modern FPGA-based FFT spectrometers provide the dynamic range and temporal resolution needed for off-line RFI detection applications.

The aim for  EBHIS  was to follow the approach presented in detail by \citet{winkel07a} which is based on a two-step procedure.
\begin{itemize}
\item Identification of spectral features. Baseline effects are minimized by a fitting procedure of the data in the time--frequency domain, e.g., using two-dimensional polynomials. Robustness of the fit is ensured by automatically setting ``windows'' around line emission or possible RFI signals.
\item Statistical and morphological considerations are applied to distinguish between astronomical emission and RFI.
\end{itemize}
These statistical detection strategies are based on the high temporal variability of interference signals while the astronomical signal needs to be considered as constant over the time interval used for the analysis. However, in order to complete a first full coverage of the Northern hemisphere we decided to scan the sky rather fast (at a rate of about $3\arcmin\,\mathrm{s}^{-1}$) which unfortunately provides only few dumps per beam size. Furthermore, some of the narrow band interference signals are very constant in time. On the other hand, practically almost all RFI events are very sharp (1--2 pixels wide) in the time--frequency plane.

\begin{figure}[!tb]
\centering
\includegraphics[scale=0.6,clip=,bb=63 84 400 216]{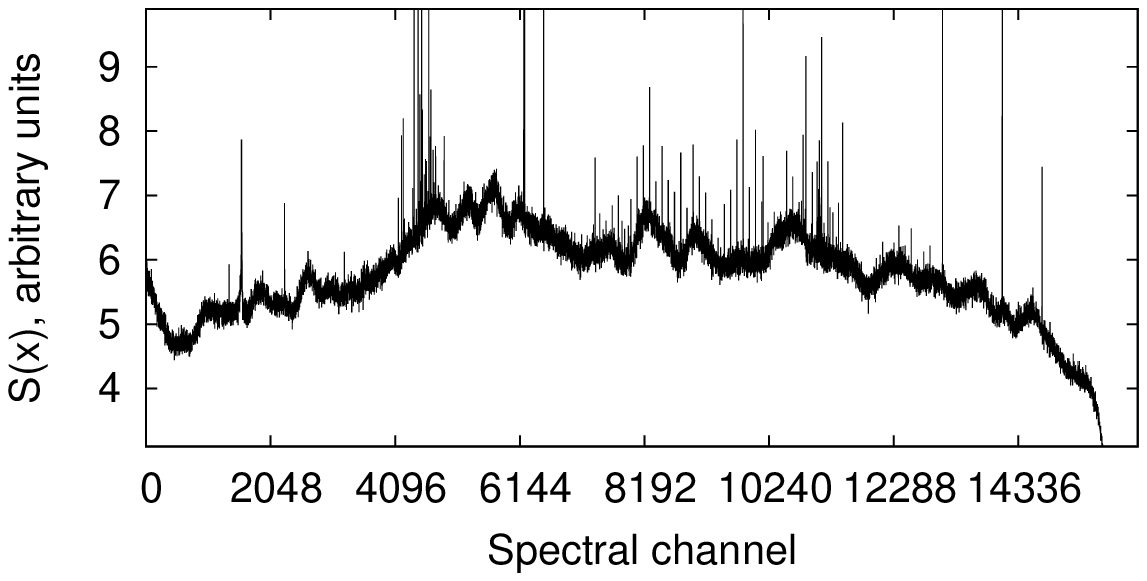}\\[0ex]
\includegraphics[scale=0.6,clip=,bb=63 48 400 216]{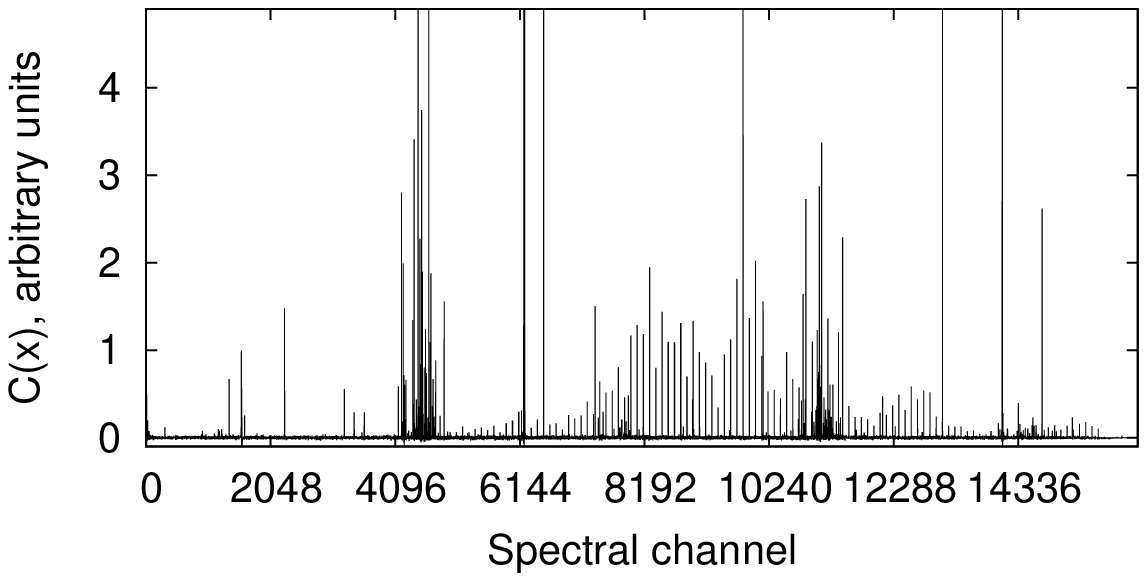}
\caption{Example spectrum (upper panel) and the applied cross-correlation with a template (lower panel) used for our RFI detection algorithm.}
\label{figcrosscorrelation}
\end{figure}

Because of that, we chose for the initial data reduction a less complex approach based on cross-correlation of the spectral data $s(x)$ with a template $p(x)$ (of size $n$),
\begin{equation}
C(x)=\sum_{i=-n/2}^{+n/2}\left( p(i)-\overline p\right)\cdot\left( s(x+i)-\overline s\right)\,.
\end{equation}
To best match the impulse-like shapes of RFI we use the templates
\begin{equation}
p(i)=\begin{cases} 1 & : i=0 \\ 0 & : i\neq0  \end{cases}\,.
\end{equation}
They are either applied to spectral data or to the time series of each spectral channel. In practice, strong RFI peaks have a large impact on $\overline s$ and $\overline p\cdot s(x)$ producing troughs in the correlation spectrum. To avoid this effect, we set  $\overline p\equiv0$ and replace the mean estimator $\overline s$ by the median, which yields a higher robustness against outliers and leads to convincing results (Figure\,\ref{figcrosscorrelation}). The correlation spectrum is searched for values in excess of more than $4\sigma$ in order to flag bad data points.

\begin{figure*}[!tb]
\centering
\includegraphics[scale=0.8,clip=,bb=66 84 344 216]{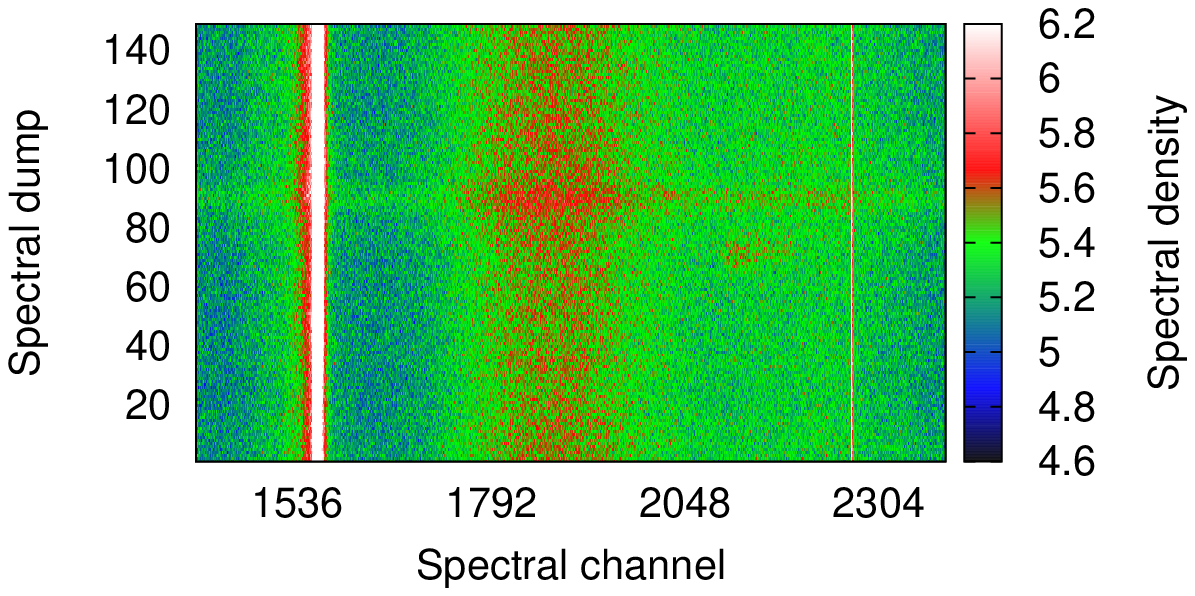}\,
\includegraphics[scale=0.8,clip=,bb=123 84 415 216]{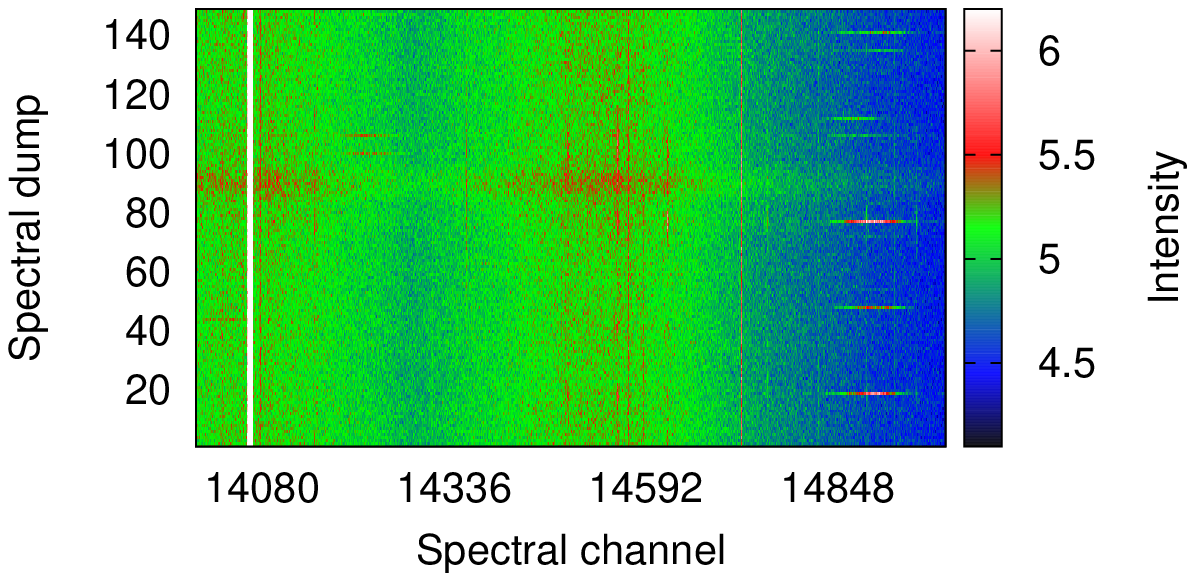}\\[0ex]
\includegraphics[scale=0.8,clip=,bb=66 49 344 216]{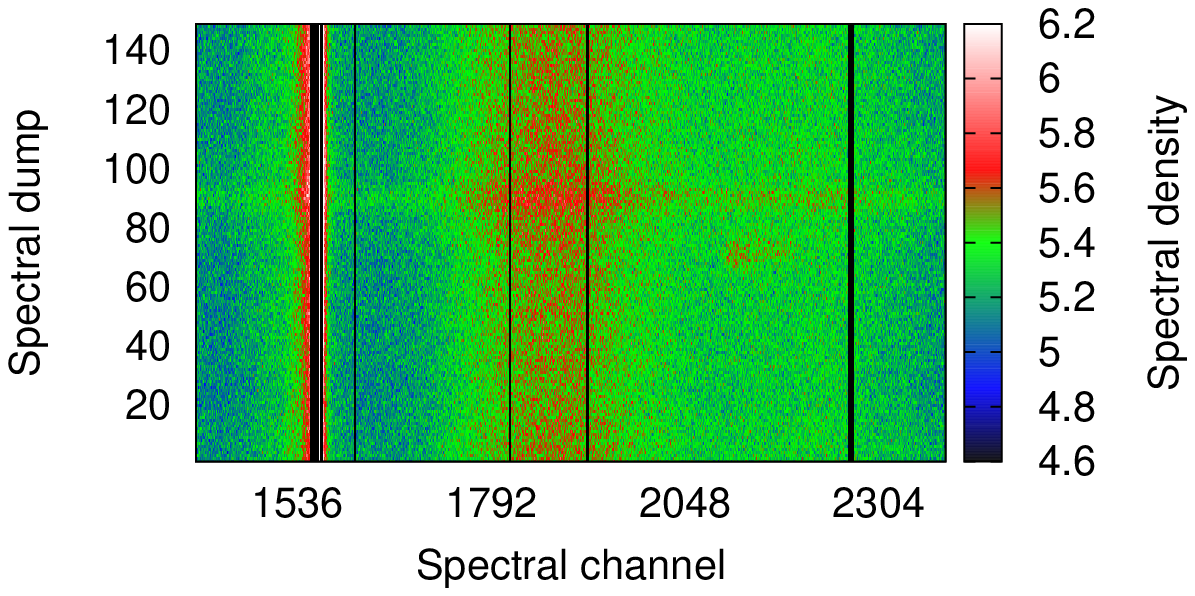}\,
\includegraphics[scale=0.8,clip=,bb=123 49 415 216]{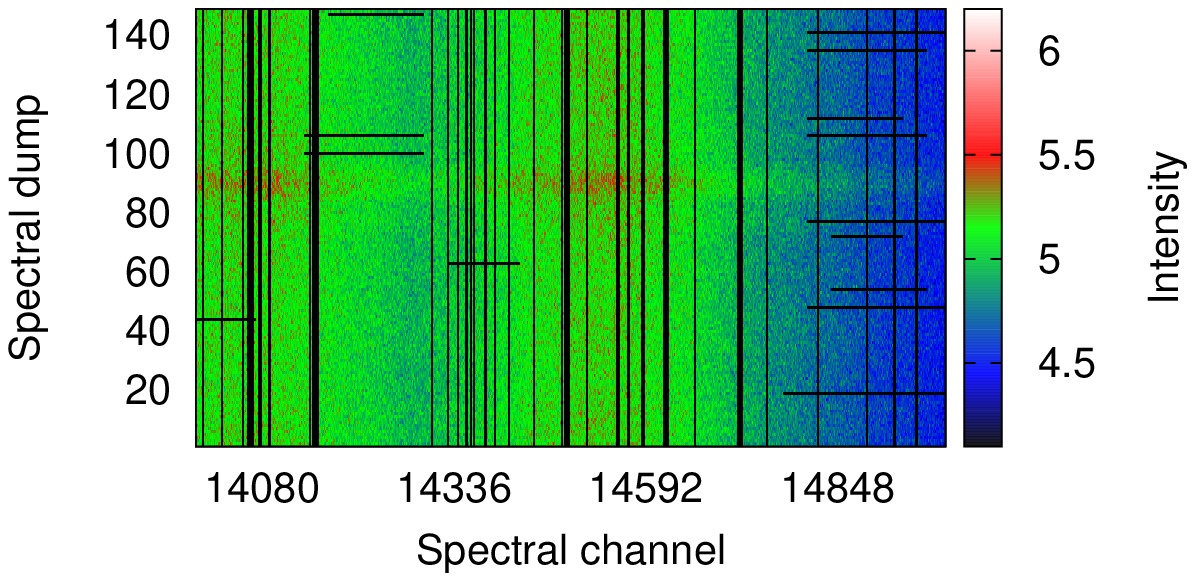}
\caption{For two distinct spectral ranges, the results of the RFI detection procedure are visualized. The upper panels contain raw spectral density plots as a function of the spectral channel and spectral dump (each dump is a 500\,ms snapshot). The lower panels show the identified RFI candidates as black areas. Note that the continuum source (around spectral dump 90) and the spectral line of a galaxy (around spectral channel 2100, dump 70) were not flagged, as desired. Sharp features of the MW emission (spectral channel $\sim1550$), however, can sometimes trigger a false detection. The algorithm also works well on broadband events and ``forests'' of narrow band RFI as shown in the bottom right panel. }
\label{figrfidetection}
\end{figure*}

Figure\,\ref{figrfidetection} shows  two regions of the time--frequency plane (upper panels) and the result of the flagging algorithm (bottom panels). The left panels contain not only a continuum source, but also a bright galaxy, neither of which produces a false detection. The right panels exhibit several broadband RFI events as well as lots of narrow-band signals which are correctly flagged by the procedure.

Note that the specific choice of our templates is suited to optimize the detection rate for the narrow (in time or frequency) RFI signals which make up for almost all events. To further improve the detection efficiency we use variable threshold levels as a function of the number of feeds and polarization channels containing the same RFI event. A more detailed explanation of the algorithms and results of simulations 
the aim of which was to quantitatively estimate the RFI detection efficiency of our software will be given in \citet{floeer10}.

\subsection{Flux calibration}\label{subsecfluxcalibration}
\subsubsection{Iterative Gaussian smoothing}
Many of the subsequent data reduction tasks rely on the computation of baselines or smoothed data for various purposes. Any RFI signals would highly distort the usual polynomial fits or Gaussian smoothing algorithms. Having an RFI flag database, contaminated data points can be flagged such that the results are not affected by these outliers. However, it is not always guaranteed that all interference signals were correctly identified and the line emission of the astronomical signals of interest would still have impact, e.g., on the baseline fits. Therefore, a more robust algorithm was developed (iterative Gaussian smoothing (IGS)) which does not rely on an RFI flag database. It works iteratively: (1) low-pass filtering the input data using a Gaussian kernel, (2) searching for outliers in the residual (i.e., peaks above a threshold of $3\sigma$ in the residual spectrum, which are flagged), and (3) replacing flagged data points in the input data with either interpolated or prior values (if available). After usually a few iterations the approximation converges.

%
%
\subsubsection{Gain curves}
The multi-beam receiver uses the heterodyne principle where the radio frequency (RF) signal is mixed with a monochromatic signal of an LO. An appropriate low-pass filter applied after this operation provides the desired IF signal at much lower carrier frequencies. The whole system can be described by
\begin{equation}
 P_\mathrm{IF}= G_\mathrm{IF}G_\mathrm{RF}  
\left [ T_\mathrm{A}+T_\mathrm{sys}\right]
\end{equation}
with $P_\mathrm{IF}$ being the power in the IF chain. $G_\mathrm{IF}$ and $G_\mathrm{RF}$ are the frequency-dependent gain functions at IF and RF stages, respectively. The gain acts on the astronomical signal of interest, $T_\mathrm{A}$, plus the contribution $T_\mathrm{sys}$ incorporating noise due to sky, ground radiation (scattering and spillover), and thermal noise of the receiver and calibration diode, etc. In our case, we also observe a rather strong multi-modal sine-wave pattern which is dependent on elevation, feed rotation angle, and polarization channel. We will discuss this issue in more detail in a subsequent paragraph.

The EBHIS data are observed using in-band frequency switching, using a frequency shift of 3\,MHz. Eventually, in the near future the receiving system can be upgraded to support the recently developed least-squares frequency switching \citep[LSFS;][]{heiles07} method for which \citet{winkel07b} proposed further improvements. LSFS requires changes to the hardware (i.e., to provide more than two LO frequencies), which were not yet introduced at the 100-m telescope. However, both methods can fail in the case of non-trivial RF gain curves, an issue which is discussed at the end of this subsection. Here, we make no specific use of the in-band frequency switching technique in order to remove gain curve effects.

To calibrate the data, recorded in arbitrary units, i.e., spectrometer counts, the product $G\equiv G_\mathrm{IF}(\nu_\mathrm{IF})G_\mathrm{RF}(\nu_\mathrm{IF})$ has to be determined. This can be done using the built-in noise diode which is fed into the system at every second spectral dump. Then
\begin{equation}
\begin{split}
 P_\mathrm{IF}^\mathrm{cal}-P_\mathrm{IF}&= G 
\left [ T^\mathrm{cal}_\mathrm{A}+T^\mathrm{cal}_\mathrm{sys}+T^\mathrm{cal}-T_\mathrm{A}-T_\mathrm{sys}\right]\\
&=GT^\mathrm{cal}
\end{split}
\end{equation}
as long as the condition $T_\mathrm{A}\simeq T^\mathrm{cal}_\mathrm{A}$ and $T_\mathrm{sys}\simeq T^\mathrm{cal}_\mathrm{sys}$ applies for adjacent spectral dumps. Furthermore, this approach assumes that the output of the noise diode is fed into the very beginning of the signal processing chain, which is not exactly true for the seven-feed receiver. For this instrument, the diodes are embedded in the waveguides connected to the feed horns. Hence, the influence of the feeds themselves is not taken into account. Furthermore, the functional form of $G$ can only be reconstructed if $T^\mathrm{cal}$ is frequency independent. Fortunately, this is indeed the case to high precision for the 21-cm seven-feed receiver (R. Keller, private communication 2009).

\begin{figure}[!t]
\centering
\includegraphics[width=0.45\textwidth,clip=,bb=63 48 423 293]{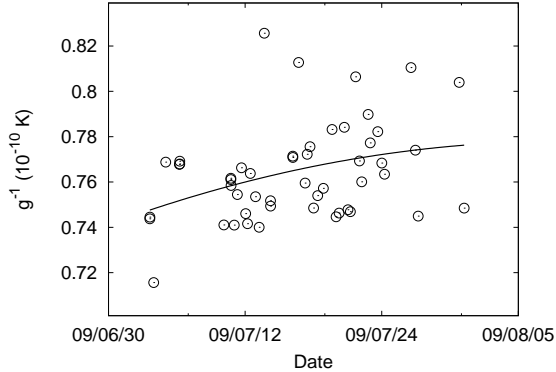}
\caption{Calculated gain factors, $g\equiv G(v_\mathrm{lsr}=0)$, obtained using the standard calibration source S\,7 over a period of about a month. There is a small overall drift, which was fitted using a third-order polynomial. Remaining scatter is about 2.5\% with few outliers, which could be flagged as bad measurements. Usually, the value of the polynomial fit is used for the gain calibration.}
\label{figs7gainfactors}
\end{figure}

The thermal noise temperature of the calibration diode, $T^\mathrm{cal}$, is known, such that one can obtain $G=(P_\mathrm{IF}^\mathrm{cal}-P_\mathrm{IF})/T^\mathrm{cal}$. However, a better precision of the absolute flux calibration can be reached by using an astronomical calibration source. For this purpose, we utilize IAU standard calibration sources (usually S\,7, because it is circumpolar for Effelsberg, sometimes also S\,8). The spectral line flux of these calibrators is well known  \citep{kalberla82} allowing one to determine the gain factor, $g\equiv G(v_\mathrm{lsr}=0)$, with an accuracy of  2\% typically. Note, that the calibrators are not continuum sources, but are well-defined \ion{H}{i} regions in the Milky Way; hence they only provide the gain for $v_\mathrm{lsr}=0$. In Figure\,\ref{figs7gainfactors},  values of $g$ for a time range of a few weeks are shown. There is a slight dependence of time which is expected for a typical receiving system. To measure the scatter about the long-time behavior a third-order polynomial was fitted, showing residual rms of about 2.5\%.
Practically, $g$ is measured by performing a polynomial fit to the calibration source spectra to separate the spectral line from the baselines and calculating
\begin{equation}
g= \frac{\sum_{v_1}^{v_2}P_\mathrm{IF}(v_\mathrm{lsr})}{S_\mathrm{cal}}\,,
\end{equation}
$v_{1,2}$ are the integration limits for the chosen IAU position according to \citet{kalberla82}.

The complete gain curve is easily computed by normalizing $GT^\mathrm{cal}$ such that $\hat G\equiv GT^\mathrm{cal}(v_\mathrm{lsr}=0\,\mathrm{km\,s}^{-1})=1$ and applying the absolute flux calibration (as obtained using the IAU standard calibration sources) by multiplying with $g$:
\begin{equation}
G=g \hat G\,.
\end{equation}

\begin{figure}[!t]
\centering
\includegraphics[scale=0.6,clip=,bb=65 84 417 167]{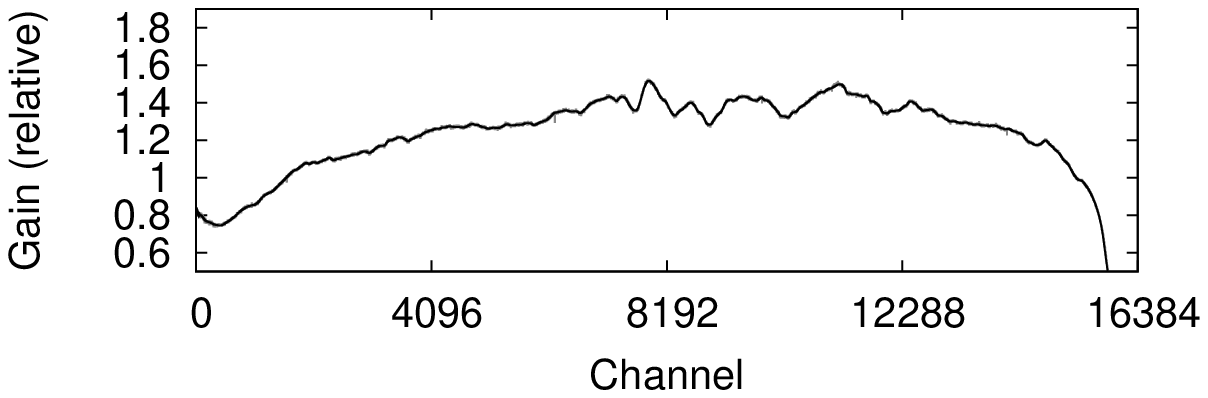}\hbox{\raise3.4em\vbox{\moveleft16.8em\hbox{\small(a)}}}\\[0ex]
\includegraphics[scale=0.6,clip=,bb=65 84 417 167]{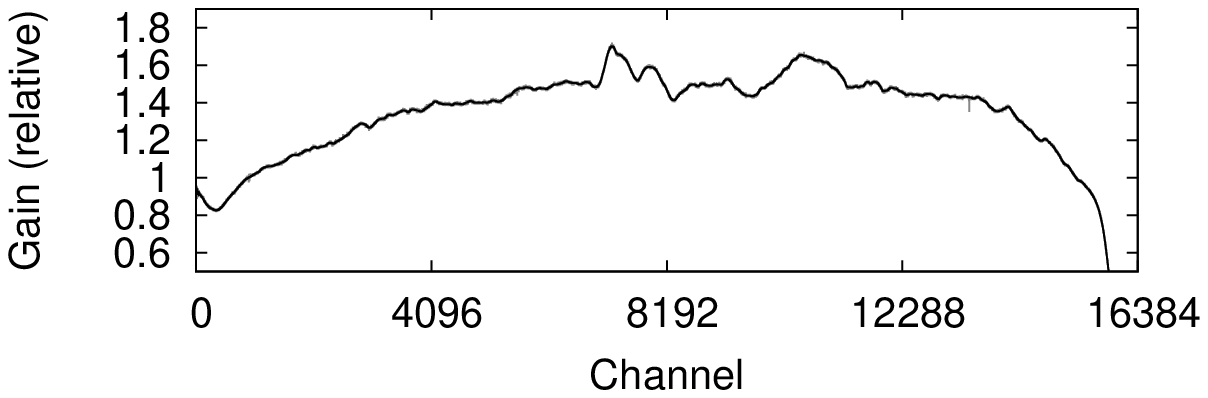}\hbox{\raise3.4em\vbox{\moveleft16.8em\hbox{\small(b)}}}\\[0ex]
\includegraphics[scale=0.6,clip=,bb=65 84 417 167]{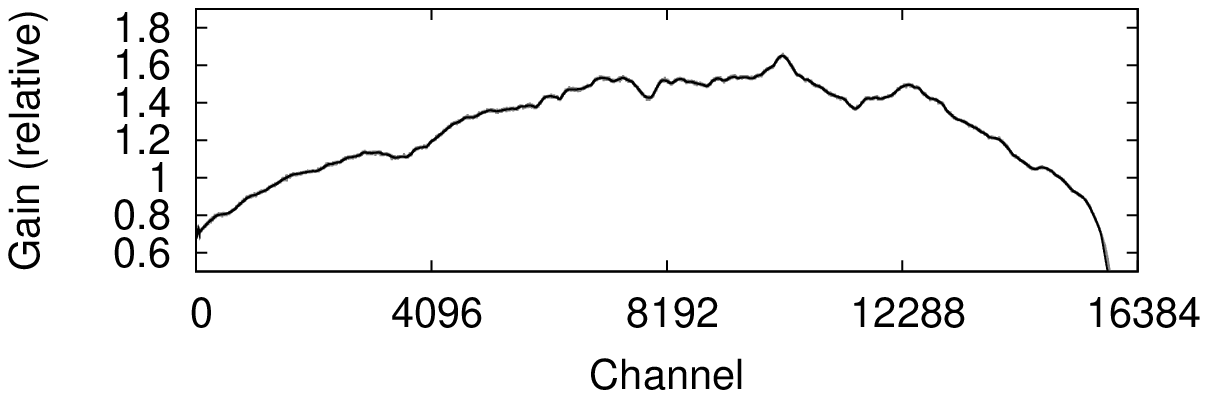}\hbox{\raise3.4em\vbox{\moveleft16.8em\hbox{\small(c)}}}\\[0ex]
\includegraphics[scale=0.6,clip=,bb=65 49 417 167]{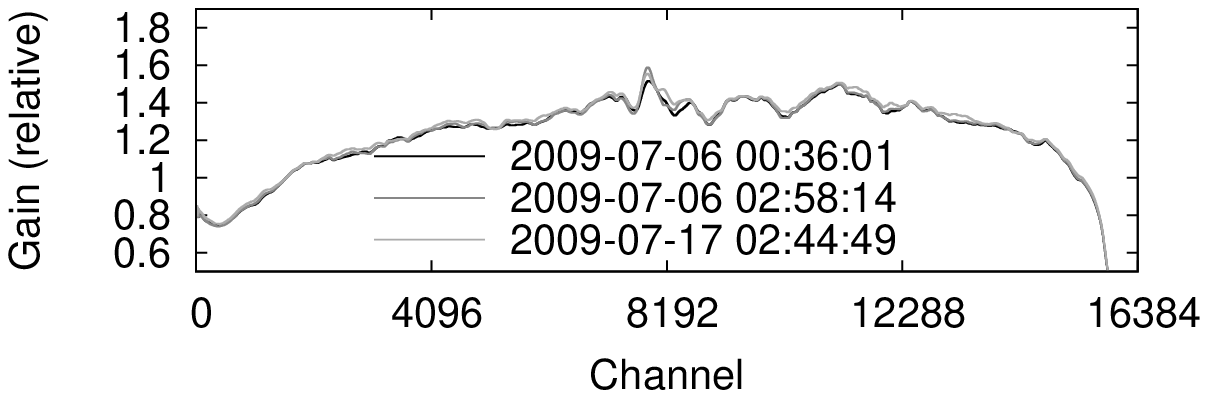}\hbox{\raise5.2em\vbox{\moveleft16.8em\hbox{\small(d)}}}
\caption{Gain curves $\hat G$ for different measurements. Panels (a) and (b) contain the left-hand polarization channel of the central feed for the two different frequency shifts. Panel (c) shows the gain curve of one of the offset feeds of the same data set. The noisy results (gray solid lines) were over plotted with a smoothed curve (black solid lines) as obtained after filtering with IGS. In panel (d), $\hat G$ (smoothed) is plotted for three different measurements (central feed) to investigate the long-term stability of the gains. The overall shape of $\hat G$ is obviously rather stable, though there are some features, especially in the center of the band, which vary more strongly.}
\label{figgaincurves}
\end{figure}

In Figure\,\ref{figgaincurves}, some of the obtained gain curves $\hat G$ are shown. They were computed by calculating the median of $\hat G$ of all spectral dumps. Usually this median gain curve contains only few outliers due to RFI signals. For later application of $\hat G$ to the spectra the gain curves are smoothed with IGS (using a filter kernel width of $\sigma=64\,\mathrm{kHz}$) to reduce residual noise. Panels (a) and (b) contain the left-hand polarization channel of the central feed for the two different frequency shifts. Several ripples are visible which follow the LO shift and, hence, can be attributed to the RF part of the gain curve. It is possible that these features are caused by resonances in the waveguide connecting the feed horn antenna with the receiver. In panel (c), a linear polarization channel of one of the offset feeds is shown. 

\begin{figure*}[!t]
\centering
\includegraphics[scale=0.65,clip=,bb=63 96 760 240]{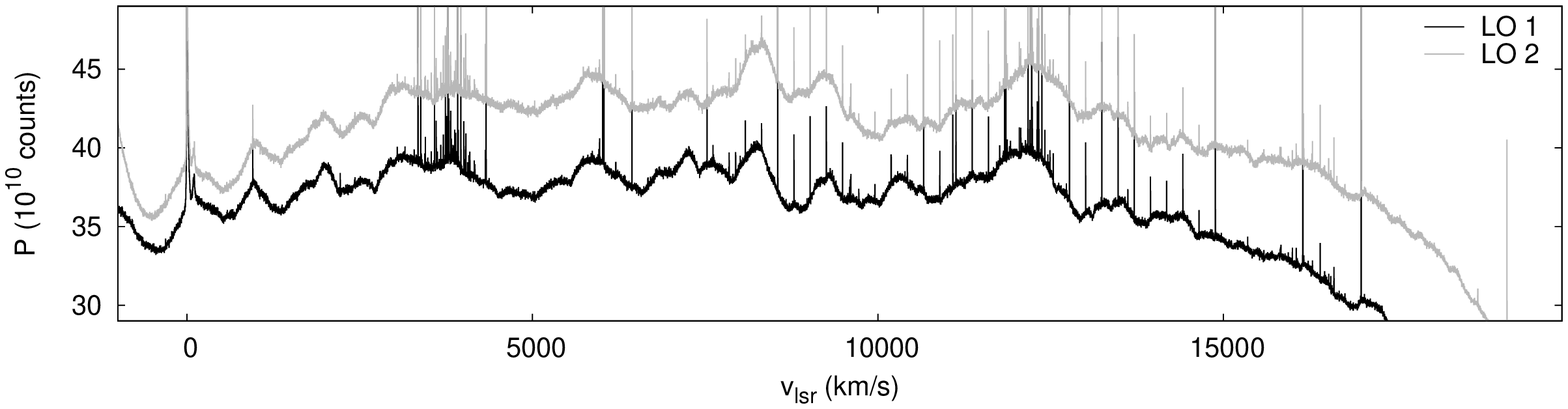}\\
\includegraphics[scale=0.65,clip=,bb=63 58 760 240]{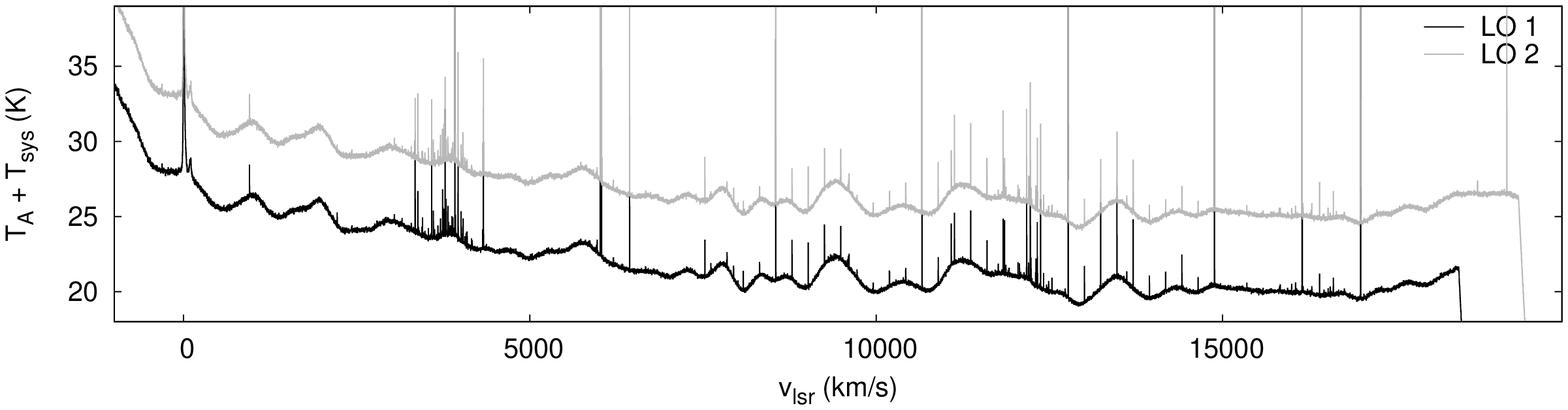}
\caption{Raw (top panel) and calibrated (bottom panel) spectra (integrated over one subscan) for the two different LO frequencies. The latter were computed by dividing the raw spectra by the calculated gain curves $\hat G$ (see Figure\,\ref{figgaincurves}) and multiplying with the absolute flux calibration values $g$. In the calibrated spectra, $T_\mathrm{A}+T_\mathrm{sys}$ is very well matching in both LO phases. The curve for LO\,2 was displaced by $\Delta T=5\,\mathrm{K}$ for better visualization. The baseline level of $T_\mathrm{A}+T_\mathrm{sys}$ determines the effective system temperatures valid for each radial velocity.}
\label{figcalibratedspectra}
\end{figure*}

Figure\,\ref{figcalibratedspectra} shows spectra (integrated over one subscan) for the two different LO frequencies before (upper panel) and after (lower panel) calibration. Features in the raw spectra do not match exactly. This is due to the frequency dependence of $G$. After applying the gain calibration, both LO setups produce nearly identical results (apart from RFI and noise contribution). The baseline level of the calibrated spectra is by definition equal to the system temperature.

The spectra shown in the lower panel of Figure\,\ref{figcalibratedspectra} also reveal multi-modal sine-wave contributions. Usually such a pattern is attributed to standing waves (SWs) between the primary and secondary focus. We tested this hypothesis by performing test measurements with the single-feed receiver (which is constructed in a similar way as the multi-feed). For technical reasons the sub-reflector at the 100-m telescope is tilted when observing with this instrument. These tests showed a significant reduction of the SW amplitudes. Furthermore, for the multi-beam receiver two distinct observations having a feed rotation angle differing by $180\degr$ but measured at similar elevation  were inspected. One could expect to find the same SW contribution but only in opposite (linearly polarized) feed horns of the hexagonal array; however, this is not the case. Therefore, we have to treat the problem in a non-analytical way, e.g., by baseline fitting or using a series of sine waves as proposed by \citet{peek08b}.

\begin{figure}[!t]
\centering
\includegraphics[scale=0.5,clip=,bb=68 83 521 291]{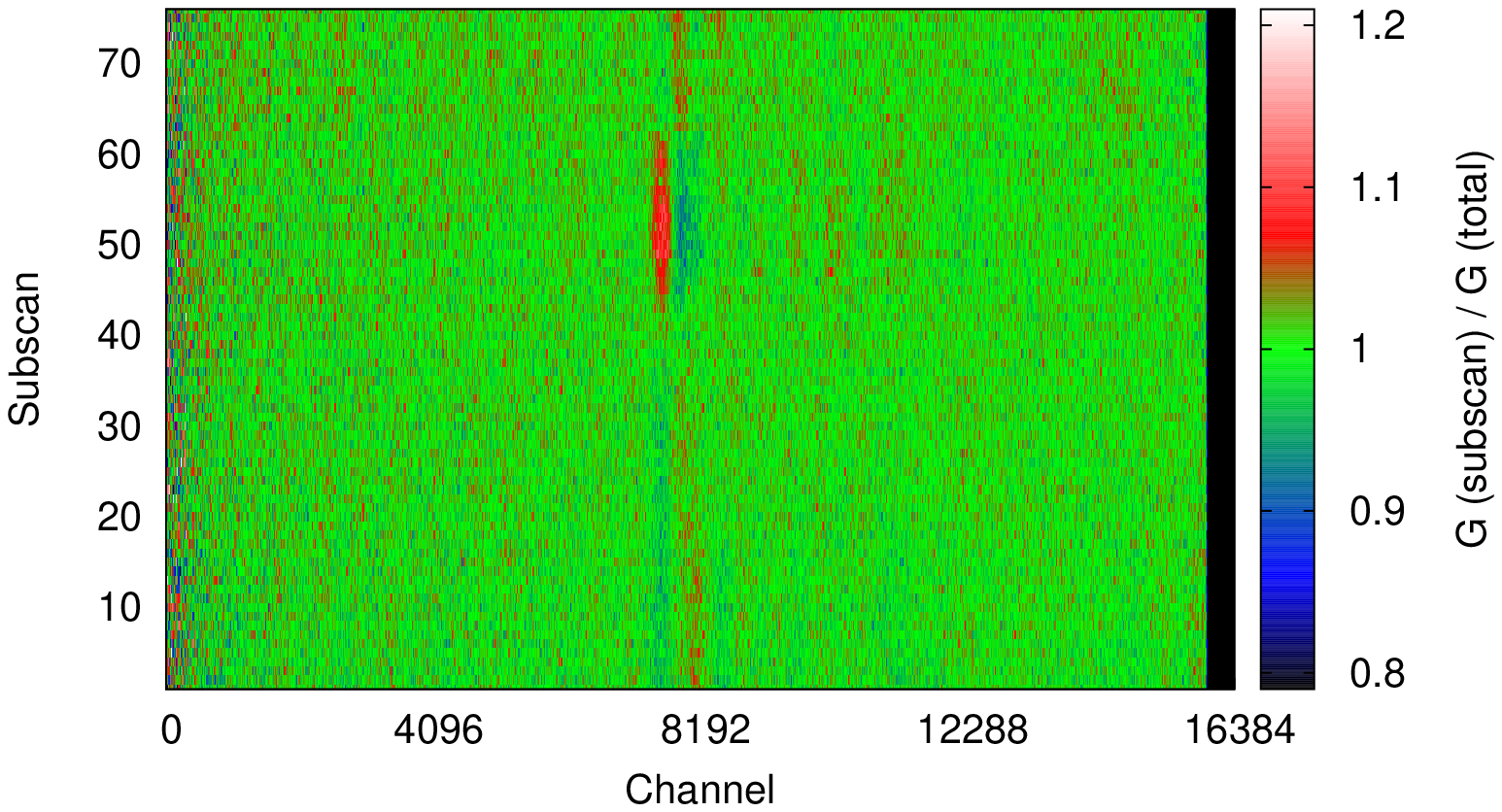}\\[0ex]
\includegraphics[scale=0.5,clip=,bb=68 48 521 291]{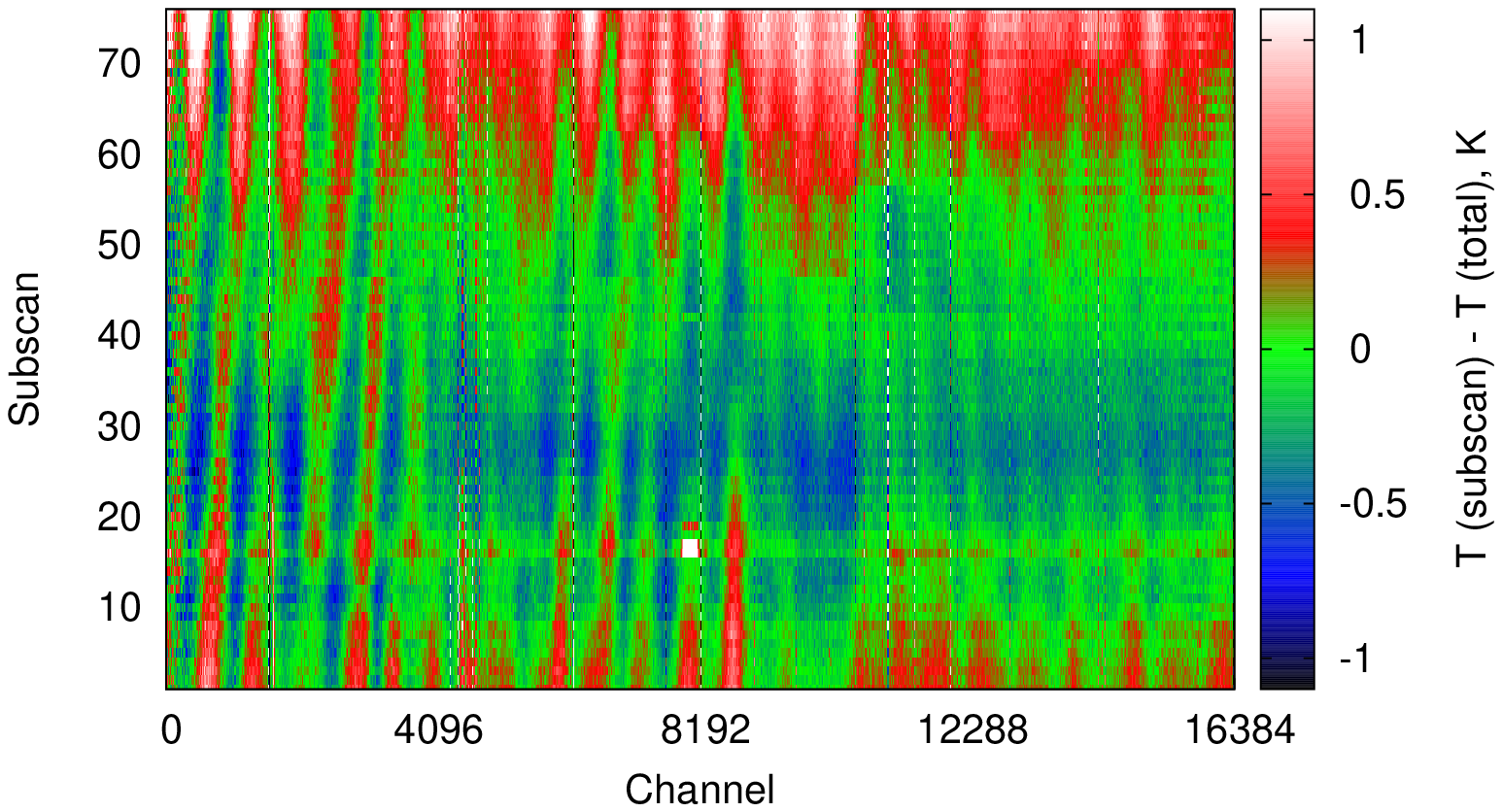}
\caption{In order to test the stability of the computed gain curves, the relative changes  of $\hat G$ are visualized in the upper panel. It shows that, except for a feature in the center of the band, the median of  $\hat G$ per subscan is equal to the total median of  $\hat G$ as calculated for the complete measurement. The bottom panel contains the relative changes of $T_\mathrm{A}+T_\mathrm{sys}$ (computed accordingly) which shows a dependency of the continuum level due to changes in elevation and a residual multi-modal sine-wave pattern which is attributed to rather fast changes in $T_\mathrm{SW}$.}
\label{figgaincurveschanges}
\end{figure}

In order to test the stability of the determined gain curves we compared $\hat G$ for three different measurements (see Figure\,\ref{figgaincurves} (d)). There are minor deviations between the observations. The relative change of the gain curve as a function of time during a single measurement is plotted in Figure\,\ref{figgaincurveschanges} (top panel). It was computed by dividing the median of $\hat G$ of the total measurement by the median of  $\hat G$ of each subscan. Except for a temporal variability in the center of the spectral band, the residuals are flat, showing that the gain curve is stable. Figure\,\ref{figgaincurveschanges} (lower panel) shows for the same measurement residual changes in $T_\mathrm{A}+T_\mathrm{sys}$ (again by computing the median for each subscan minus the total median). Two effects are obvious. First, there is an overall dependence on the elevation angle, which is a function of the subscan number, causing the continuum level to change. Second, $T_\mathrm{SW}$ obviously contains a contribution which changes slowly, likely as a function of the feed rotation angle and elevation. To test both dependencies, a measurement without any feed rotation was examined. Here, SW residuals of almost the same amplitude occur. This seems to rule out the feed angle correspondence, but the SWs are polarized \citep[see also][]{heiles05}; hence, a feed rotation should lead to a net effect. In most of our measurements the differential feed rotation is rather small such that the SWs are not significantly affected. Furthermore, several drift scans having constant elevation angles and no feed rotation were analyzed. In most cases no residual SWs show up. Few measurements contain a residual pattern, however, not such a clear sine-wave-like signal as in the lower panel of Figure\,\ref{figgaincurveschanges}. These patterns might be due to the solar irradiation, as they seem to occur only during daytime. 

Finally, we discuss a method which enables us to measure the gain calibration factors for all seven feeds simultaneously. In \citet{kalberla82}, only the flux for the exact positions on the calibration sources was determined. In order to calibrate the offset feeds, it would hence be necessary to point each of them subsequently to the exact location of the calibration region. To efficiently use the observing time (about an order of magnitude), we chose to map the area around the standard calibration sources S\,7 and S\,8. From these maps we extract reference flux values (integrated spectral line fluxes within a certain velocity interval) to calculate the gain factors for all feeds. In this scheme, only the central feed is positioned on S\,7/S\,8 exactly. Note that our calibration method uses the integral values of the feed response function rather than the peak values.

\subsubsection{Frequency switching techniques}
Using standard in-band frequency switching one usually computes the term
\begin{equation}
\begin{split}
&\frac{ P^\mathrm{sig}(f_\mathrm{IF})-P^\mathrm{ref}(f_\mathrm{IF}) }{P^\mathrm{ref}(f_\mathrm{IF})} = 
\left[ T_\mathrm{A}^\mathrm{sig}(f_\mathrm{RF}) +\left( T_\mathrm{sys}^\mathrm{sig}(f_\mathrm{RF}) - T_\mathrm{sys}^\mathrm{ref}(f_\mathrm{RF}) \right)\right. \\
&+\left.\frac{\Delta G}{G} \left( T^\mathrm{sig}_\mathrm{A}+T^\mathrm{sig}_\mathrm{sys}+T_\mathrm{A}^\mathrm{sig}(f_\mathrm{RF}) \right)\right] \times 
\left[ \frac{1- \frac{T^\mathrm{sig}_\mathrm{sys}(f_\mathrm{RF})}{T^\mathrm{sig}_\mathrm{A}+T^\mathrm{sig}_\mathrm{sys}} }{ T^\mathrm{sig}_\mathrm{A}+T^\mathrm{sig}_\mathrm{sys}} \right]
\end{split}
\end{equation}
to get rid of the gain curve \citep{heiles07}. Here, the frequency-dependent and independent (continuum) parts were treated separately. However, as \citet{heiles07} points out, the method relies highly on the assumption that the relative difference between both LO phases, $\Delta G/G\lll1$,  which is definitely not valid for the multi-beam receiver. Furthermore, the term $T_\mathrm{sys}^\mathrm{sig}(f_\mathrm{RF}) - T_\mathrm{sys}^\mathrm{ref}(f_\mathrm{RF})$ must be negligible which is not the case due to the standing wave contribution. \citet{heiles05} reports about similar issues for the Arecibo telescope. Note, that the least-squares frequency switching technique is not affected by the SW contribution, but would still fail in the presence of a non-trivial RF filter curve \citep{heiles07}.

The SW problem can also have an impact on data observed by position switching 
when $T_\mathrm{sys}^\mathrm{on}(f_\mathrm{RF})\neq T_\mathrm{sys}^\mathrm{off}(f_\mathrm{RF})$, which can happen if the feed rotation or elevation angles change between both phases. Furthermore, in the vicinity of moderate or strong continuum sources, residual effects can occur. Also, if the reference spectrum is obtained by computing the median of a subscan, the elevation should not change during the scan line; otherwise the SW contribution cannot be considered to be constant.

\subsection{Stray-radiation correction}\label{subsecstrayradiationcorrection}
In order to compute the true brightness temperature, the antenna temperature $T_\mathrm{A}$ must be deconvolved to account for the sidelobe pattern of the antenna response function. This so-called SR correction, $T_\mathrm{SR}$, can be expressed as additive term such that
\begin{equation}
T_\mathrm{B}=\frac{T_\mathrm{A}}{\eta_\mathrm{eff}}-T_\mathrm{SR}
\end{equation}
with the antenna gain efficiency, $\eta_\mathrm{eff}$. SR can provide a significant fraction to the measured spectrum especially in galactic science where Milky Way emission is observed toward all lines of sight.

For EBHIS we make use of the existing SR code developed by \citet{kalberla78} which was successfully applied to the LAB \citep{kalberla05} and the GASS \citep{kalberla10} surveys. 

\subsection{Baseline fitting}\label{subsecbpcalibration}
\begin{figure*}[!t]
\centering
\includegraphics[scale=0.75,clip=,bb=67 78 688 190]{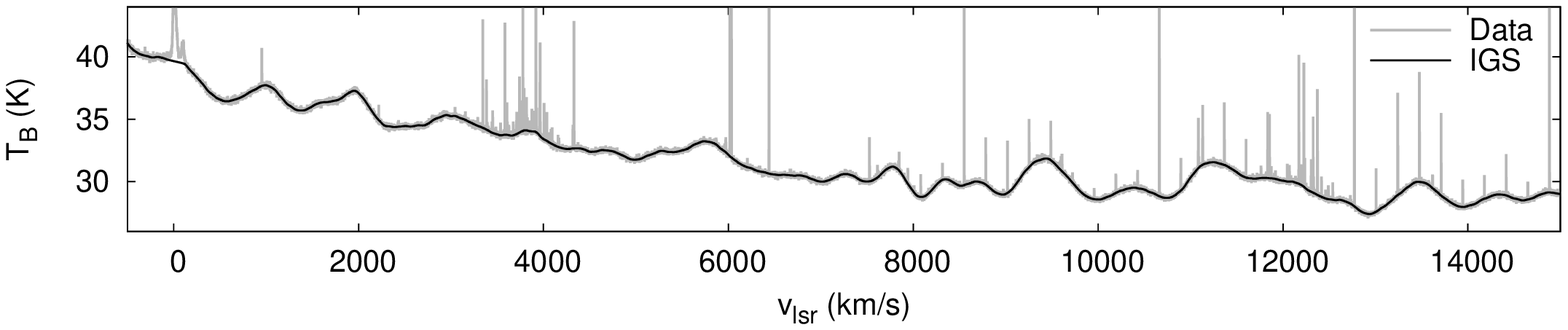}\\
\includegraphics[scale=0.75,clip=,bb=67 56 400 190]{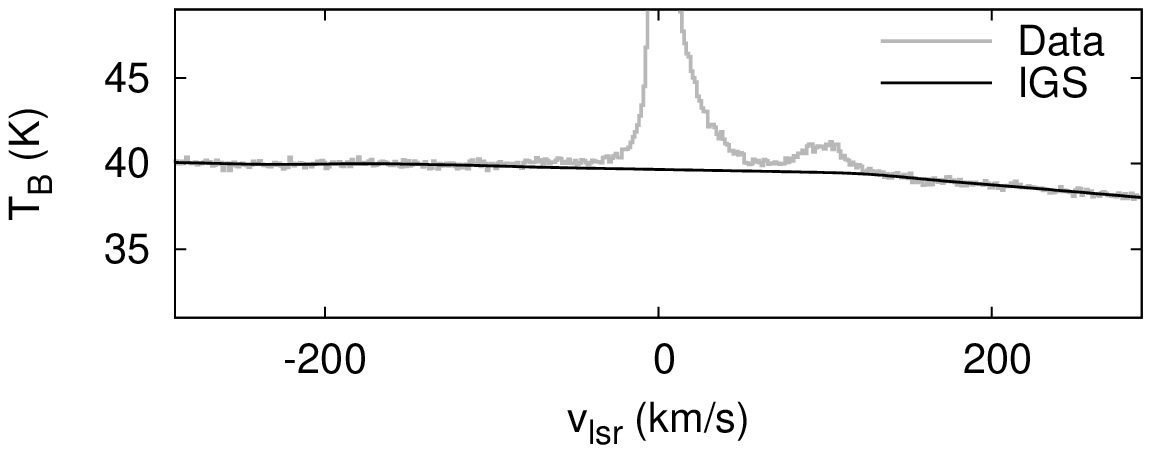}\hspace{-0.2em}
\includegraphics[scale=0.75,clip=,bb=115 56 400 190]{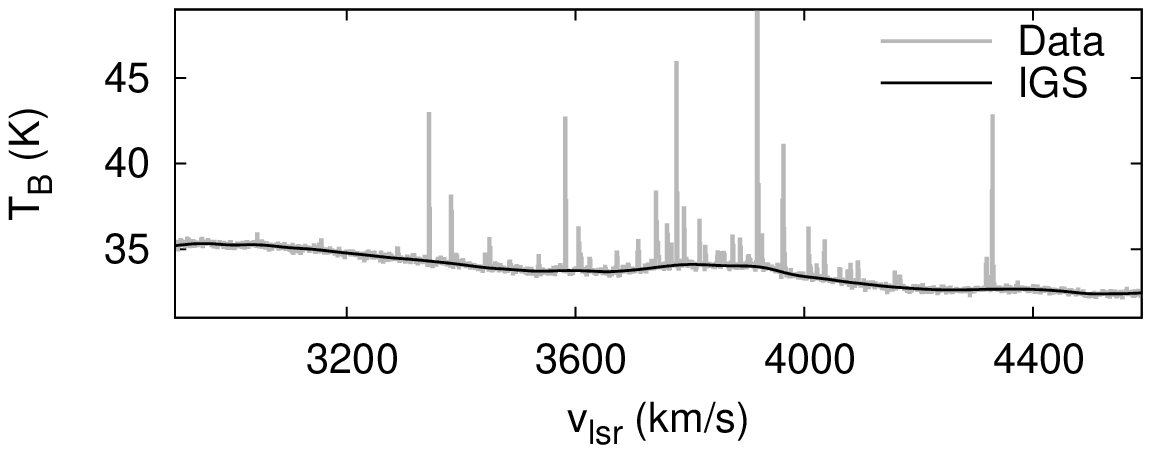}
\caption{To remove the baselines, the IGS algorithm is applied. The upper panel displays a complete spectrum, while the lower panels contain a zoom-in, showing that the filtering procedure works well in the case of the presence of strong and broad emission lines, as well as in the vicinity of RFI signals and even for extended ranges of contaminated data.}
\label{figbaselinefitting}
\end{figure*}

After flux calibration and subtraction  of the SR contribution remaining baselines, i.e., the standing waves, must be removed. Usual polynomial baseline fitting is not very well suited for our problem, as the SW contribution would need either extremely high polynomial orders or require piece-wise fitting of smaller spectral areas leading to problems of connecting the overlap regions. Therefore, we simply apply our IGS algorithm (kernel width of $\sigma=0.1\,\mathrm{MHz}$). For higher accuracy, the software allows us to perform the IGS on the median of all spectral dumps of a subscan, which in most cases leads to better results, as the baseline is only slowly changing with time. In some cases the broad diffuse part of the Milky Way line emission is not flagged to the full width. For improvement, we produced an SQL table containing spectral windows based on the LAB survey, which can be used as priors. In Figure\,\ref{figbaselinefitting}, we show an example spectrum with the baseline fit as computed for the median of all spectral dumps within a subscan.

\subsection{Gridding}\label{subsecgridding}
A lot of effort went into the development of a new gridding software, to allow serial data processing. This is necessary, as hundreds of gigabytes of spectral data have to be gridded together to form even relatively small data cubes.  Due to the serial approach only the size of the produced data cube puts constraints on the memory usage of the software, such that large areas of the sky can be processed for the full spectral range in one cycle. 

\subsubsection{Map calculation}
The contribution of each individual spectral dump to the pixels on the grid is calculated using a Gaussian of size $\sigma_\mathrm{kernel}$ as weighting function. 
The weighting factors for each pixel are accumulated in a weight map (or in a weight cube, if flags are to be applied). Finally, each pixel in the data cube is divided by the associated cumulated weighting factor. The gridding method is equivalent to a (spatial) convolution of the individual spectra with a Gaussian kernel. Consequently, the effective angular resolution is slightly degraded; typically we use a kernel width of $5\fmin 4$ (FWHM) which enlarges the effective telescope beam of $9\arcmin$ to $10\fmin 5$. 

The algorithm is independent of the chosen projection system (for the pixel grid), as true angular distances are used for calculations. Accordingly it is necessary to convert between true coordinates and the pixel representation. For such transformations, the gridder uses the world coordinate system (WCS) enhancement of the CFITSIO library\footnote{\url{http://heasarc.nasa.gov/docs/software/fitsio/fitsio.html}} \citep[WCSlib;][]{greisen02,calabretta02,greisen06} to allow gridding into many different projection systems.

The software  allows us to subtract a low-order polynomial from each dump to remove the continuum flux, which could also be used to calculate continuum maps, though the elevation-dependent contribution would need some extra treatment.

\subsubsection{Velocity regridding}
The LO frequencies at the 100-m telescope are fixed to a constant value at the beginning of each subscan (calculated for the central feed). Therefore, the spectra have to be regridded in the velocity/frequency space to account for shifts in the local standard of rest frame during a single subscan, as well as for differences between the central and the offset feeds. The  spectral regridding is performed for each individual dump before gridding to the data cube, using the Akima spline algorithm provided by the GNU scientific library\footnote{\url{http://www.gnu.org/software/gsl/}} \citep[GSL;][]{gsl}.

\section{First results}\label{secfirstresults}

\begin{figure*}[!p]
\centering
\includegraphics[scale=0.36,clip=,bb=27 185 566 625]{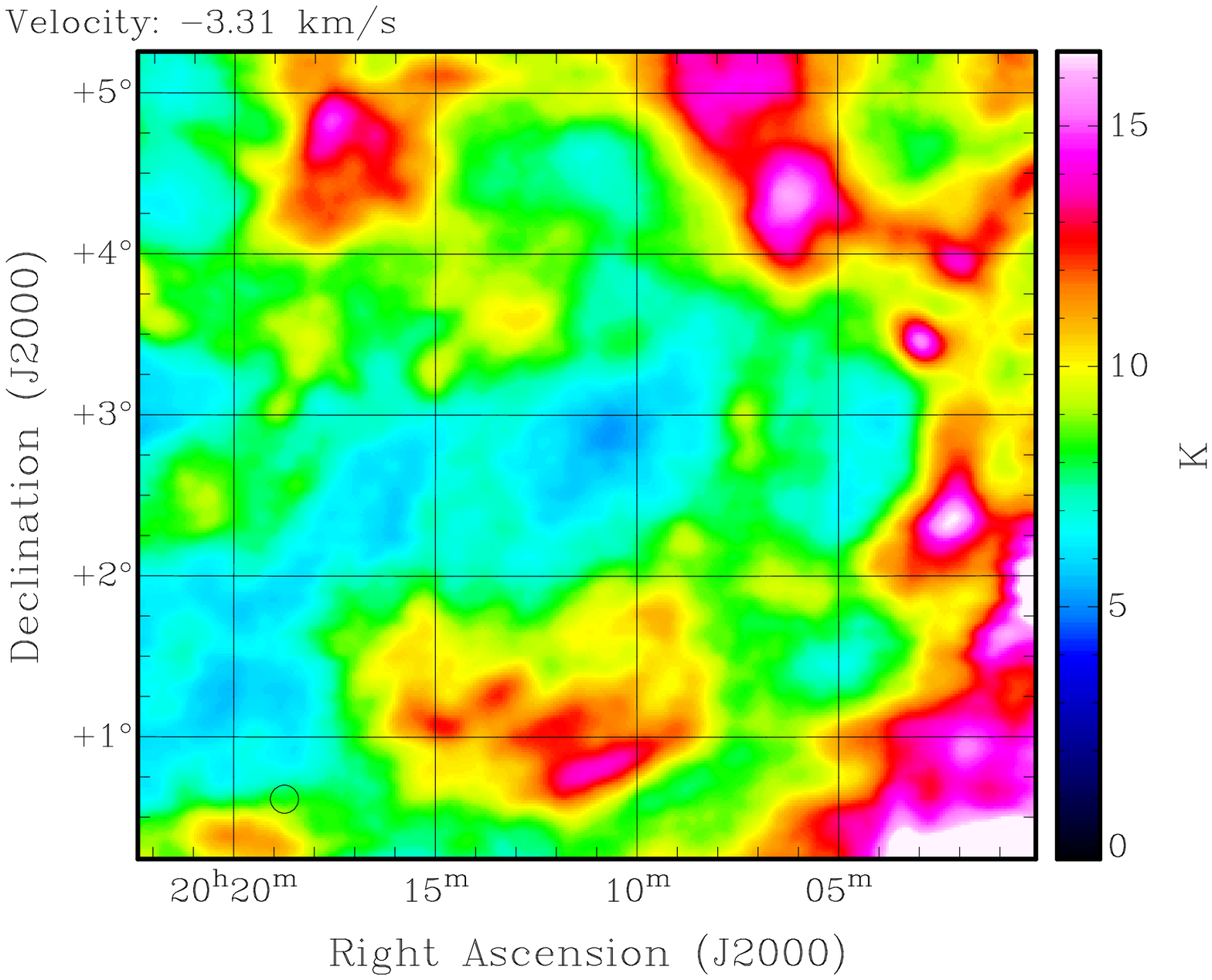}\qquad
\includegraphics[scale=0.36,clip=,bb=27 185 566 625]{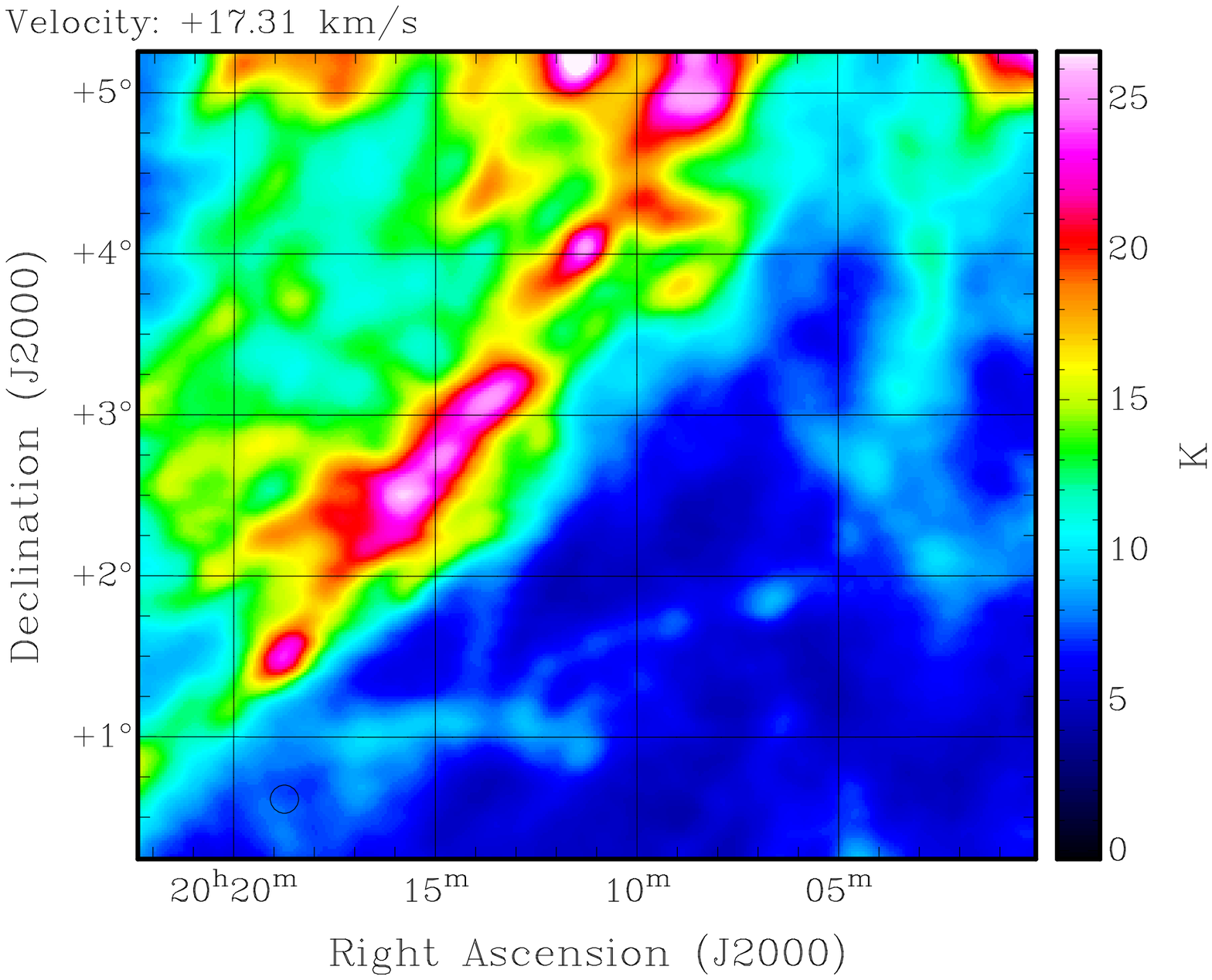}\\[0ex]
\includegraphics[scale=0.36,clip=,bb=27 185 566 625]{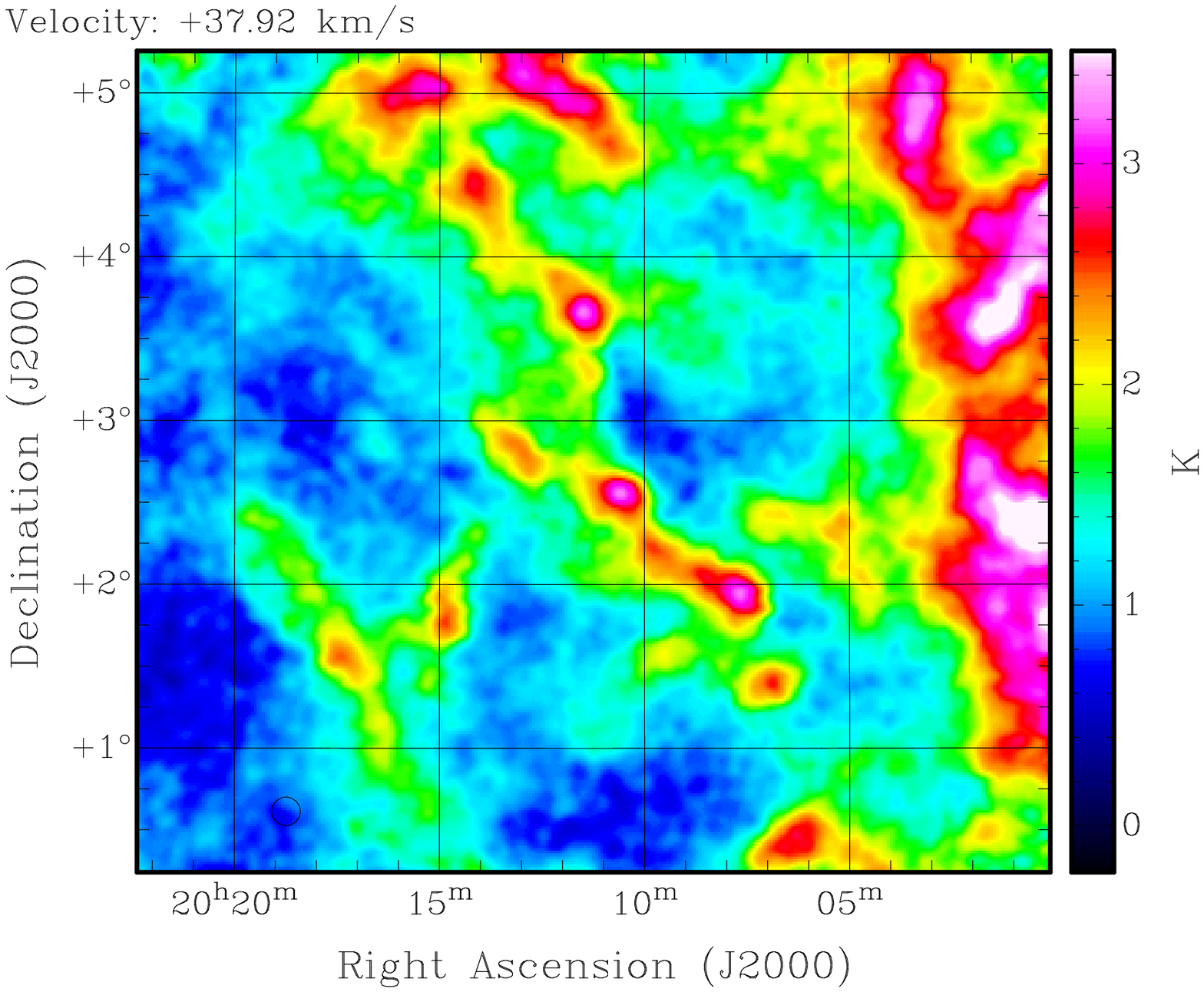}\qquad
\includegraphics[scale=0.36,clip=,bb=27 185 566 625]{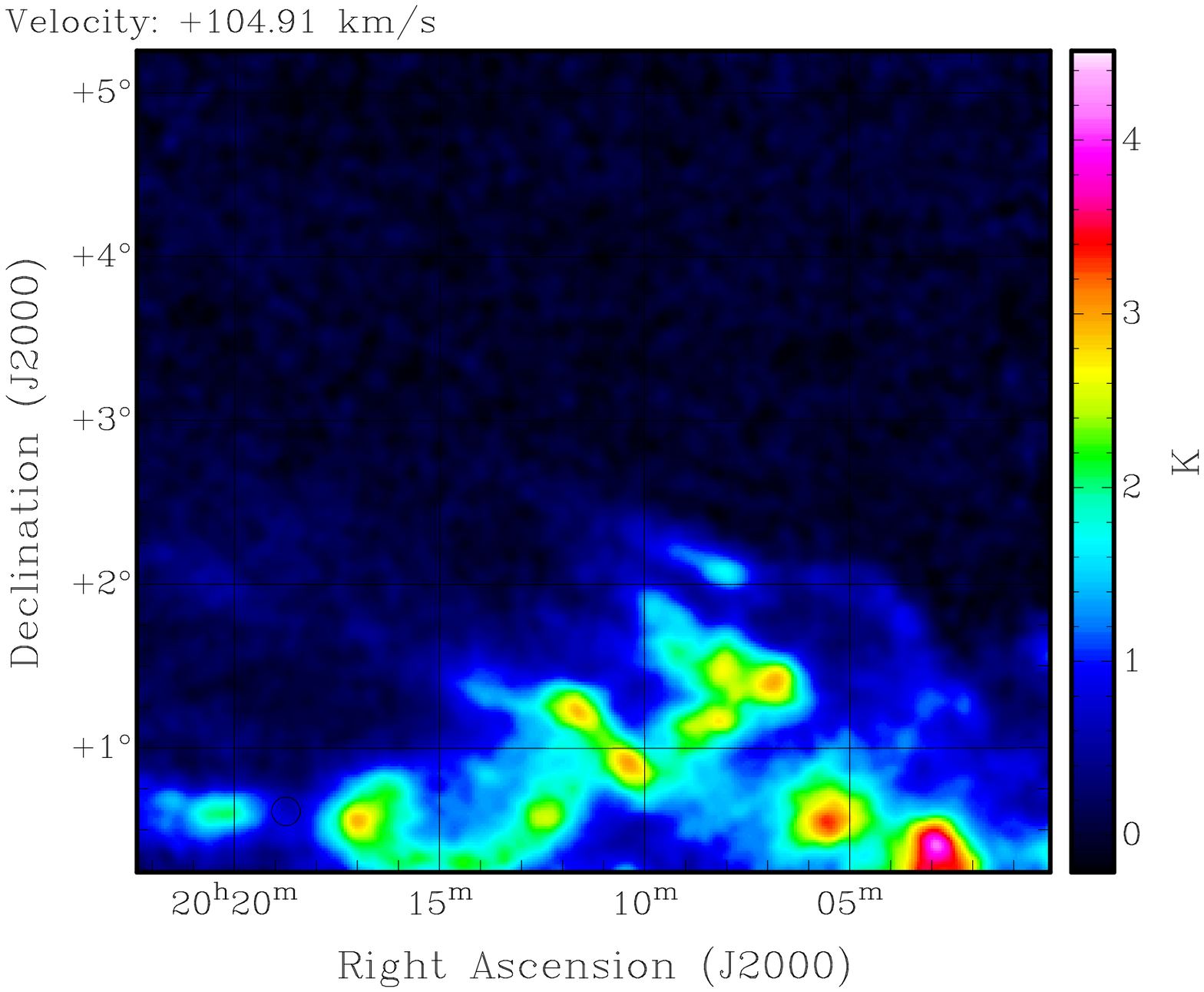}\\[2ex]
\includegraphics[width=\textwidth,clip=,bb=14 14 784 270]{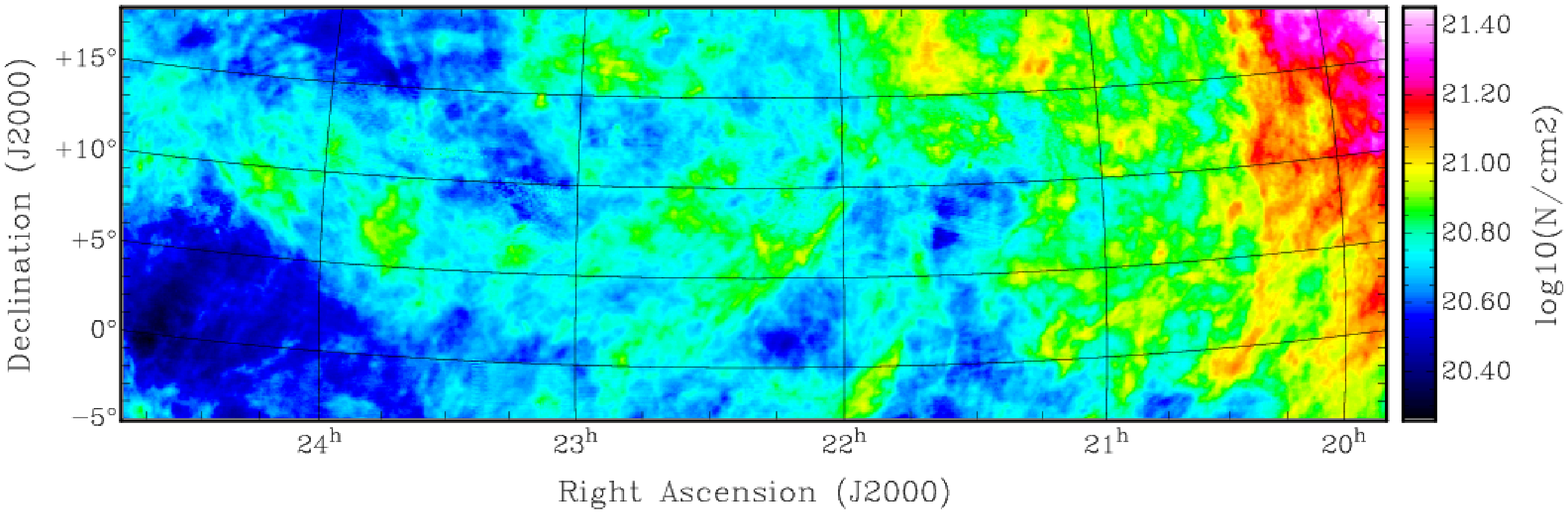}
\caption{Top and middle panels show several velocity planes of an example data cube. Lots of filaments and clumps are visible which are often unresolved by the telescope beam of $9\arcmin$ (marked by a circle in the lower left part of each map). The bottom panel contains a column density map of one of the two larger 2000\,deg$^2$ areas observed so far. $N_\mathrm{HI}$ is calculated for the velocity interval between $-50\leq v_\mathrm{lsr}\leq+50\,\mathrm{km\,s}^{-1}$.}
\label{fig0810}
\end{figure*}

\begin{figure}[!t]
\centering
\includegraphics[height=0.45\textwidth,clip=,angle=-90]{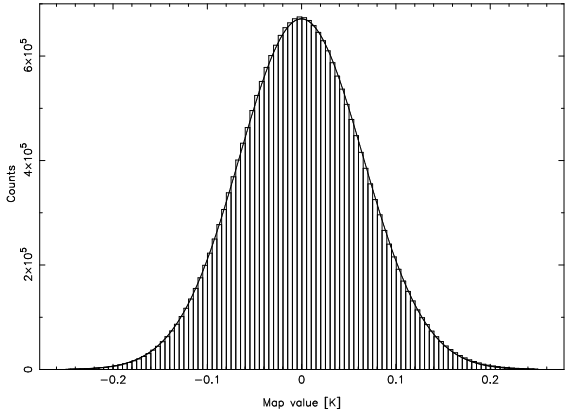}
\caption{Noise histogram for a line-free part of the data cube as calculated using 22 million pixels. The solid line shows a Gaussian fitted to the data. The resulting rms is 65\,mK which matches the theoretical expectation.}
\label{fignoisehistogram}
\end{figure}

In Figure\,\ref{fig0810}, we show as an example some velocity planes of one of our measurements showing the wealth of details which can be found in the EBHIS data. We find in this particular field a lot of IVC and HVC gas, exhibiting filaments and clumps many of which are probably not resolved with the Effelsberg telescope beam of $9\arcmin$. 
The $5\times5$ deg$^2$ map was observed for a total integration time of about 140\,minutes leading to an rms noise level of 65\,mK, equivalent to a $4\sigma$ column density limit of about $3.6\cdot10^{18}\,\mathrm{cm}^{-2}$ (calculated for a Gaussian-shaped emission line of width $20\,\mathrm{km\,s}^{-1}$ and spectral data smoothed to $10\,\mathrm{km\,s}^{-1}$, i.e., a $4\sigma$ detection in two adjacent spectral channels). These values match the theoretical expectation (due to the low elevation angle, the system temperature was relatively high with 40\,K). For a region of the data cube being free of any emission the miriad task \texttt{imhist} was used to calculate a noise histogram (see Figure\,\ref{fignoisehistogram}). It reveals a distribution which is almost perfectly described by a Gaussian. Remaining RFI, which would show up as deviations from the standard distribution, is not visible.

To achieve a first full coverage of the northern sky expected for Spring 2011, the scan speed was increased by a factor of 2. Therefore, we expect for the first data release a noise level of $\lesssim90\,\mathrm{mK}$ (or $5\cdot10^{18}\,\mathrm{cm}^{-2}$) which is very similar to the noise level of the GASS survey (when smoothing EBHIS data to the GASS beam size). 

Currently, about a third of the northern hemisphere has been observed so far. The coverage is inhomogeneously distributed over many smaller test fields (1 to 100\,deg$^2$) and two larger coherent sky portions, each covering about 2000\,deg$^2$. The first large area is toward the northern galactic pole and the second one toward the northern tip of the Magellanic Leading Arm. In Figure\,\ref{fig0810} (bottom panel), we show a column density map of the latter calculated for the velocity interval between $-50\leq v_\mathrm{lsr}\leq+50\,\mathrm{km\,s}^{-1}$. 

\begin{figure*}[!p]
\centering
\includegraphics[scale=0.45,clip=,bb=55 245 536 610]{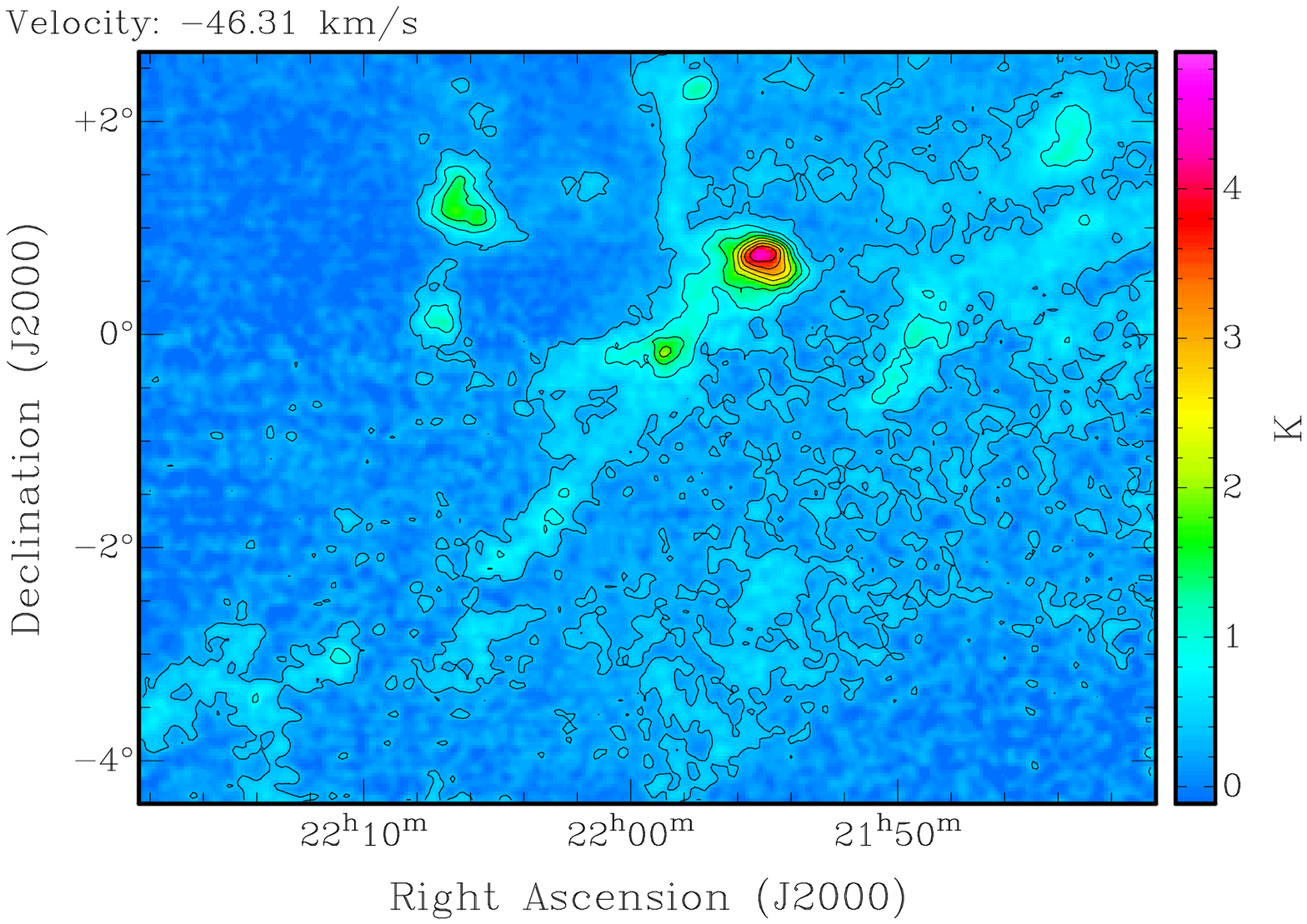}\hbox{\raise15.2em\vbox{\moveleft18.3em\hbox{EBHIS}}}
\includegraphics[scale=0.45,clip=,bb=111 245 595 610]{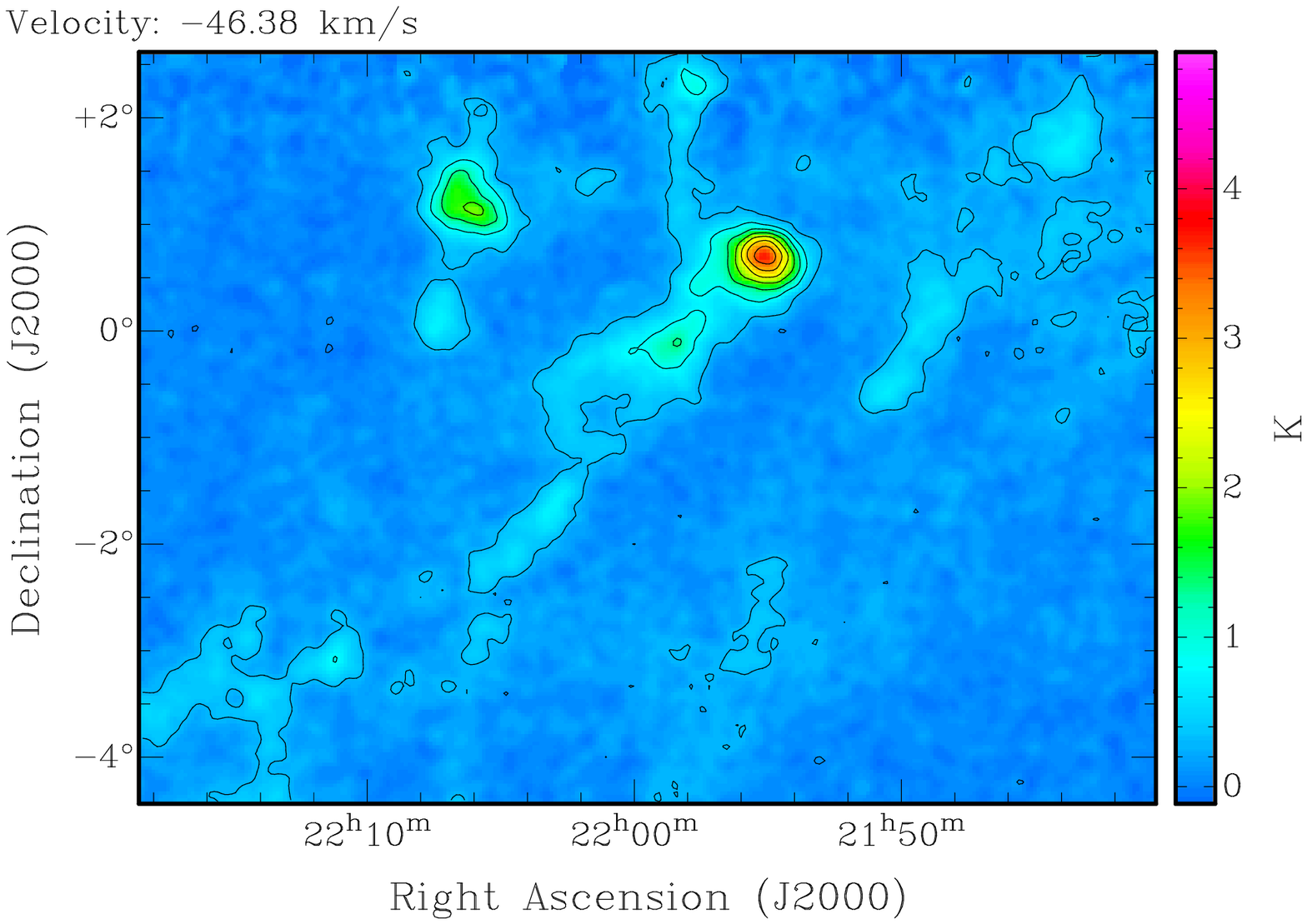}\hbox{\raise15.2em\vbox{\moveleft20.5em\hbox{GASS}}}\\[1ex]
\includegraphics[width=0.7\textwidth,clip=,bb=14 14 683 514]{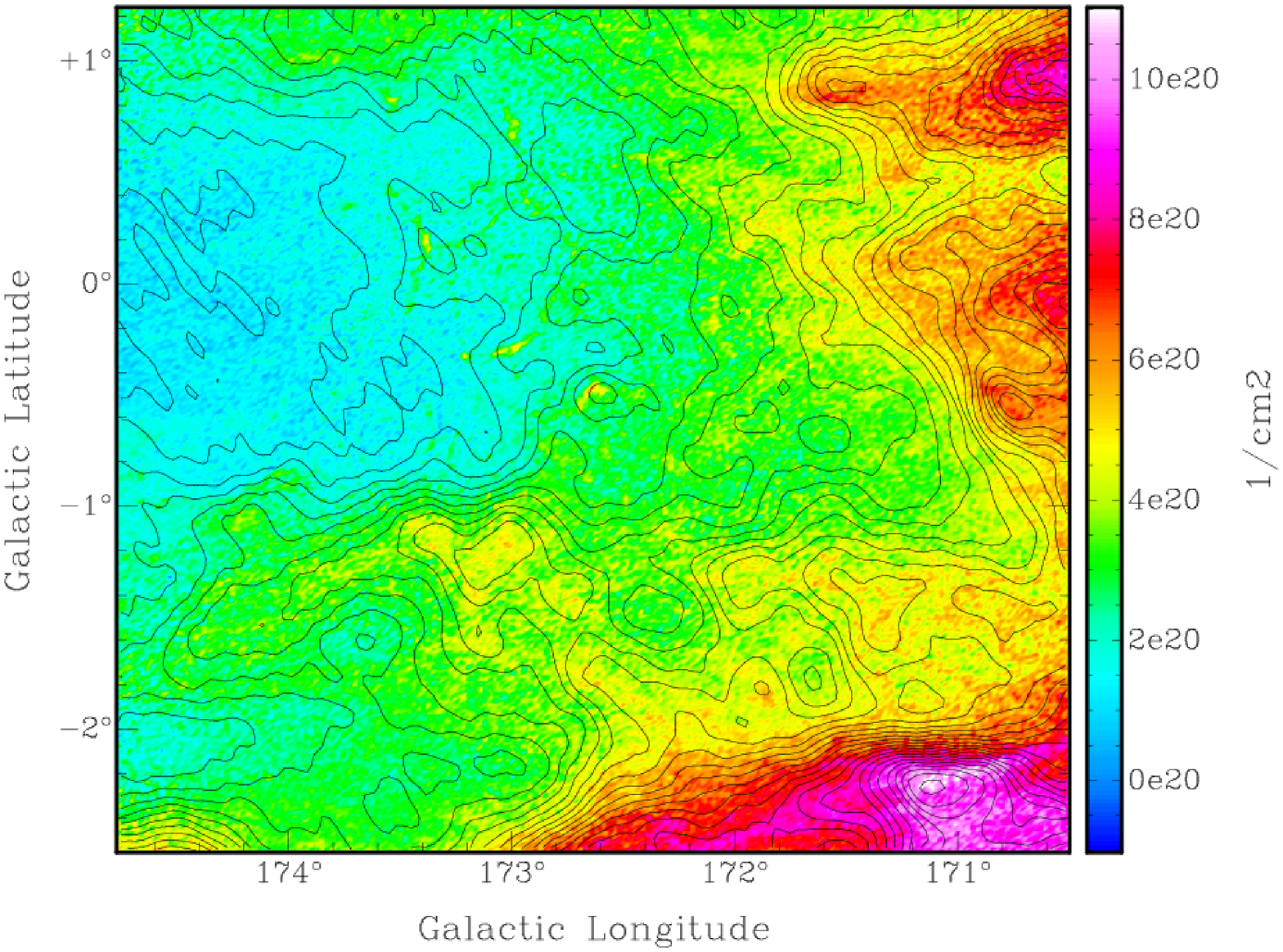}\hbox{\raise22.9em\vbox{\moveleft28.3em\hbox{CGPS, EBHIS}}}\\[-2ex]
\caption{Comparison of the EBHIS data with other surveys. The top panels contain channel maps at $v_\mathrm{lsr}=-46\,\mathrm{km\,s}^{-1}$ from  EBHIS (left) and GASS (right). Note  that the EBHIS map is actually a mosaic of four different ($5\times5$\,deg$^2$) observations where the measurements above  declination $0\udeg$ had two times higher integration time than the southern fields. The rms noise for EBHIS is about 90\,mK (65\,mK in the northern part), GASS has overall about 60\,mK. The contour lines start at 0.3\,K and increase in steps of 0.5\,K. The bottom panel shows column densities (integrated over $-27\leq v_\mathrm{lsr}\leq-46\,\mathrm{km\,s}^{-1}$) from the CGPS superposed with EBHIS data marked by contours, ranging from $1\cdot10^{20}\,\mathrm{cm}^{-2}$ to $8.5\cdot10^{20}\,\mathrm{cm}^{-2}$ in steps of $2.5\cdot10^{19}\,\mathrm{cm}^{-2}$. }
\label{figoverlaps}
\end{figure*}

To evaluate the EBHIS data in comparison with previous surveys, we compare in Figure\,\ref{figoverlaps} EBHIS with GASS and a selected area from the Canadian Galactic Plane Survey \citep[CGPS;][]{taylor03}. The upper panel displays channel maps of EBHIS and GASS ($v_\mathrm{lsr}=-46\,\mathrm{km\,s}^{-1}$) showing an intermediate-velocity structure. The contour levels are the same for both data sets, though the nominal noise levels are different. Due to the higher angular resolution of EBHIS more details are visible. Figure\,\ref{figoverlaps} (bottom panel) shows an overlap region of the CGPS superposed with EBHIS contour lines. Despite the different angular resolution, EBHIS traces the same structures as the CGPS.

\section{Summary}\label{secsummary}
We presented in this work the current status of the EBHIS data reduction procedures. The observed spectral data contain lots of RFI and exhibit a rather complicated gain curve shape. A lot of effort went into the investigation of these issues and the development of robust and computationally fast algorithms. We discussed the standing wave problem and its potential impact on the widely used frequency or position switching techniques. 

The receiver quality was extensively tested. System temperatures are as expected and the complete receiving system has stability times sufficiently high to reach all scientific aims of the EBHIS. Using the standard calibrators S\,7 or S\,8, a gain calibration accuracy of better than 3\% is easily obtained. 

The current quality of the EBHIS data and the reduction software is promising. We are confident that even the first data release in 2011 will provide a database in quality comparable to HIPASS and GASS. This will also allow us to produce a great successor to the LAB survey data by combining EBHIS and GASS both having much higher sensitivity and resolution.

\acknowledgments

\textit{Acknowledgments.} 
The authors thank the Deutsche Forschungsgemeinschaft (DFG) for financial support under the research grant KE757/7-1. The shown results are based on observations with the 100-m telescope of the MPIfR (Max-Planck-Institut f\"ur Radioastronomie) at Effelsberg. Our work would not have been possible without the continuous support from the MPIfR and Effelsberg staff. This research used the facilities of the Canadian Astronomy Data Centre operated by the National Research Council of Canada with the support of the Canadian Space Agency. The research presented in this paper has used data from the Canadian Galactic Plane Survey a Canadian project with international partners supported by the Natural Sciences and Engineering Research Council.






\bibliographystyle{aa}
\bibliography{references}

\begin{thebibliography}{32}
\expandafter\ifx\csname natexlab\endcsname\relax\def\natexlab#1{#1}\fi

\bibitem[{{Adelman-McCarthy} {et~al.}(2008){Adelman-McCarthy}, {Ag{\"u}eros},
  {Allam}, {Allende Prieto}, {Anderson}, {Anderson}, {Annis}, {Bahcall},
  {Bailer-Jones}, {Baldry}, {Barentine}, {Bassett}, {Becker}, {Beers}, {Bell},
  {Berlind}, {Bernardi}, {Blanton}, {Bochanski}, {Boroski}, {Brinchmann},
  {Brinkmann}, {Brunner}, {Budav{\'a}ri}, {Carliles}, {Carr}, {Castander},
  {Cinabro}, {Cool}, {Covey}, {Csabai}, {Cunha}, {Davenport}, {Dilday}, {Doi},
  {Eisenstein}, {Evans}, {Fan}, {Finkbeiner}, {Friedman}, {Frieman},
  {Fukugita}, {G{\"a}nsicke}, {Gates}, {Gillespie}, {Glazebrook}, {Gray},
  {Grebel}, {Gunn}, {Gurbani}, {Hall}, {Harding}, {Harvanek}, {Hawley},
  {Hayes}, {Heckman}, {Hendry}, {Hindsley}, {Hirata}, {Hogan}, {Hogg}, {Hyde},
  {Ichikawa}, {Ivezi{\'c}}, {Jester}, {Johnson}, {Jorgensen}, {Juri{\'c}},
  {Kent}, {Kessler}, {Kleinman}, {Knapp}, {Kron}, {Krzesinski}, {Kuropatkin},
  {Lamb}, {Lampeitl}, {Lebedeva}, {Lee}, {Leger}, {L{\'e}pine}, {Lima}, {Lin},
  {Long}, {Loomis}, {Loveday}, {Lupton}, {Malanushenko}, {Malanushenko},
  {Mandelbaum}, {Margon}, {Marriner}, {Mart{\'{\i}}nez-Delgado}, {Matsubara},
  {McGehee}, {McKay}, {Meiksin}, {Morrison}, {Munn}, {Nakajima}, {Neilsen},
  {Newberg}, {Nichol}, {Nicinski}, {Nieto-Santisteban}, {Nitta}, {Okamura},
  {Owen}, {Oyaizu}, {Padmanabhan}, {Pan}, {Park}, {Peoples}, {Pier}, {Pope},
  {Purger}, {Raddick}, {Re Fiorentin}, {Richards}, {Richmond}, {Riess}, {Rix},
  {Rockosi}, {Sako}, {Schlegel}, {Schneider}, {Schreiber}, {Schwope}, {Seljak},
  {Sesar}, {Sheldon}, {Shimasaku}, {Sivarani}, {Smith}, {Snedden}, {Steinmetz},
  {Strauss}, {SubbaRao}, {Suto}, {Szalay}, {Szapudi}, {Szkody}, {Tegmark},
  {Thakar}, {Tremonti}, {Tucker}, {Uomoto}, {Vanden Berk}, {Vandenberg},
  {Vidrih}, {Vogeley}, {Voges}, {Vogt}, {Wadadekar}, {Weinberg}, {West},
  {White}, {Wilhite}, {Yanny}, {Yocum}, {York}, {Zehavi}, \&
  {Zucker}}]{adelman08}
{Adelman-McCarthy}, J.~K., {Ag{\"u}eros}, M.~A., {Allam}, S.~S., {et~al.} 2008,
  \apjs, 175, 297

\bibitem[{{Barnes} {et~al.}(2001){Barnes}, {Staveley-Smith}, {de Blok},
  {Oosterloo}, {Stewart}, {Wright}, {Banks}, {Bhathal}, {Boyce}, {Calabretta},
  {Disney}, {Drinkwater}, {Ekers}, {Freeman}, {Gibson}, {Green}, {Haynes}, {te
  Lintel Hekkert}, {Henning}, {Jerjen}, {Juraszek}, {Kesteven}, {Kilborn},
  {Knezek}, {Koribalski}, {Kraan-Korteweg}, {Malin}, {Marquarding}, {Minchin},
  {Mould}, {Price}, {Putman}, {Ryder}, {Sadler}, {Schr{\"o}der}, {Stootman},
  {Webster}, {Wilson}, \& {Ye}}]{barnes01}
{Barnes}, D.~G., {Staveley-Smith}, L., {de Blok}, W.~J.~G., {et~al.} 2001,
  \mnras, 322, 486

\bibitem[{{Bhat} {et~al.}(2005){Bhat}, {Cordes}, {Chatterjee}, \&
  {Lazio}}]{bhat05}
{Bhat}, N.~D.~R., {Cordes}, J.~M., {Chatterjee}, S., \& {Lazio}, T.~J.~W. 2005,
  Radio Science, 40, 5

\bibitem[{{Bradley} \& {Barnbaum}(1996)}]{bradleybarnbaum96}
{Bradley}, R. \& {Barnbaum}, C. 1996, in Bulletin of the American Astronomical
  Society, Vol.~28, 1418

\bibitem[{{Briggs} {et~al.}(2000){Briggs}, {Bell}, \& {Kesteven}}]{briggs00}
{Briggs}, F.~H., {Bell}, J.~F., \& {Kesteven}, M.~J. 2000, \aj, 120, 3351

\bibitem[{{Calabretta} \& {Greisen}(2002)}]{calabretta02}
{Calabretta}, M.~R. \& {Greisen}, E.~W. 2002, \aap, 395, 1077

\bibitem[{{Fl\"{o}er} {et~al.}(2010){Fl\"{o}er}, {Winkel}, \&
  {Kerp}}]{floeer10}
{Fl\"{o}er}, L., {Winkel}, B., \& {Kerp}, J. 2010, in Proc. RFI2010, RFI
  Mitigation Workshop, Groningen, 2010 March 29--31, ed. W.~{Baan}, A.-J.
  {Boonstra}, \& M.~{Lewis}, Proceedings of Science, PoS(RFI2010)042

\bibitem[{{Fridman}(2001)}]{fridman01}
{Fridman}, P.~A. 2001, \aap, 368, 369

\bibitem[{{Galassi} {et~al.}(2009){Galassi}, {Davies}, {Theiler}, {Gough},
  {Jungman}, {Alken}, {Booth}, \& {Rossi}}]{gsl}
{Galassi}, M., {Davies}, J., {Theiler}, J., {et~al.} 2009, {GNU Scientific
  Library Reference Manual - Third Edition} (Network Theory Ltd)

\bibitem[{{Giovanelli} {et~al.}(2005){Giovanelli}, {Haynes}, {Kent},
  {Perillat}, {Saintonge}, {Brosch}, {Catinella}, {Hoffman}, {Stierwalt},
  {Spekkens}, {Lerner}, {Masters}, {Momjian}, {Rosenberg}, {Springob},
  {Boselli}, {Charmandaris}, {Darling}, {Davies}, {Garcia Lambas}, {Gavazzi},
  {Giovanardi}, {Hardy}, {Hunt}, {Iovino}, {Karachentsev}, {Karachentseva},
  {Koopmann}, {Marinoni}, {Minchin}, {Muller}, {Putman}, {Pantoja}, {Salzer},
  {Scodeggio}, {Skillman}, {Solanes}, {Valotto}, {van Driel}, \& {van
  Zee}}]{giovanelli05p}
{Giovanelli}, R., {Haynes}, M.~P., {Kent}, B.~R., {et~al.} 2005, \aj, 130, 2598

\bibitem[{{Goldsmith}(2004)}]{goldsmith04}
{Goldsmith}, P.~F. 2004, in Bulletin of the American Astronomical Society,
  Vol.~36, 1475

\bibitem[{{Greisen} \& {Calabretta}(2002)}]{greisen02}
{Greisen}, E.~W. \& {Calabretta}, M.~R. 2002, \aap, 395, 1061

\bibitem[{{Greisen} {et~al.}(2006){Greisen}, {Calabretta}, {Valdes}, \&
  {Allen}}]{greisen06}
{Greisen}, E.~W., {Calabretta}, M.~R., {Valdes}, F.~G., \& {Allen}, S.~L. 2006,
  \aap, 446, 747

\bibitem[{{Heiles}(2005)}]{heiles05}
{Heiles}, C. 2005, {Fixed Pattern Noise}, Tech. rep., National Astronomy and
  Ionosphere Center (Arecibo: NAIC)

\bibitem[{{Heiles}(2007)}]{heiles07}
{Heiles}, C. 2007, \pasp, 119, 643

\bibitem[{{Heiles} {et~al.}(2004){Heiles}, {Goldston}, {Mock}, {Parsons},
  {Stanimirovic}, \& {Werthimer}}]{heiles04}
{Heiles}, C., {Goldston}, J., {Mock}, J., {et~al.} 2004, in Bulletin of the
  American Astronomical Society, Vol.~36, 1476

\bibitem[{{Kalberla}(1978)}]{kalberla78}
{Kalberla}, P.~M.~W. 1978, PhD thesis, University of Bonn

\bibitem[{{Kalberla} {et~al.}(2005){Kalberla}, {Burton}, {Hartmann}, {Arnal},
  {Bajaja}, {Morras}, \& {P{\"o}ppel}}]{kalberla05}
{Kalberla}, P.~M.~W., {Burton}, W.~B., {Hartmann}, D., {et~al.} 2005, \aap,
  440, 775

\bibitem[{{Kalberla} \& {Kerp}(2009)}]{kalberla09}
{Kalberla}, P.~M.~W. \& {Kerp}, J. 2009, \araa, 47, 27

\bibitem[{{Kalberla} {et~al.}(2010){Kalberla}, {McClure-Griffiths}, {Pisano},
  {Calabretta}, {Ford}, {Lockman}, {Staveley-Smith}, {Kerp}, {Winkel},
  {Murphy}, {Nakanishi}, \& {Newton-McGee}}]{kalberla10}
{Kalberla}, P.~M.~W., {McClure-Griffiths}, N.~M., {Pisano}, D.~J., {et~al.}
  2010, submitted

\bibitem[{{Kalberla} {et~al.}(1982){Kalberla}, {Mebold}, \&
  {Reif}}]{kalberla82}
{Kalberla}, P.~M.~W., {Mebold}, U., \& {Reif}, K. 1982, \aap, 106, 190

\bibitem[{{Keller} {et~al.}(2006){Keller}, {Nalbach}, {M\"{u}ller}, {Teuber},
  {Sch\"{a}fer}, {Klein}, {Kr\"{a}mer}, {Bell}, \&
  {Meyer}}]{keller06techreport}
{Keller}, R., {Nalbach}, M., {M\"{u}ller}, K., {et~al.} 2006, Multi-Beam
  Receiver for Beam-Park Experiments and Data Collection Unit for Beam Park
  Experiments with Multi-Beam Receivers, Tech. rep., Max-Planck-Institut
  f\"{u}r Radioastronomie, (Bonn: MPIfR)

\bibitem[{{Klein} {et~al.}(2008){Klein}, {Kr{\"a}mer}, {Hochg{\"u}rtel},
  {G{\"u}sten}, {Bell}, {Meyer}, \& {Chetik}}]{klein08}
{Klein}, B., {Kr{\"a}mer}, I., {Hochg{\"u}rtel}, S., {et~al.} 2008, in
  Proceedings of the Ninteenth International Symposium on Space Terahertz
  Technology, held April 28-30, 2008, in Groningen. Edited by Wolfgang Wild.
  Sponsored by SRON, Netherlands Institute for Space Research, TUDelft, Delft
  University of Technology, and the University of Groningen. 2008, p.192, ed.
  {W.~Wild}, 192--+

\bibitem[{{Klein} {et~al.}(2006){Klein}, {Philipp}, {Kr{\"a}mer}, {Kasemann},
  {G{\"u}sten}, \& {Menten}}]{klein06}
{Klein}, B., {Philipp}, S.~D., {Kr{\"a}mer}, I., {et~al.} 2006, \aap, 454, L29

\bibitem[{{McClure-Griffiths} {et~al.}(2006){McClure-Griffiths}, {Ford},
  {Pisano}, {Gibson}, {Staveley-Smith}, {Calabretta}, {Dedes}, \&
  {Kalberla}}]{mcclure06}
{McClure-Griffiths}, N.~M., {Ford}, A., {Pisano}, D.~J., {et~al.} 2006, \apj,
  638, 196

\bibitem[{{McClure-Griffiths} {et~al.}(2009){McClure-Griffiths}, {Pisano},
  {Calabretta}, {Ford}, {Lockman}, {Staveley-Smith}, {Kalberla}, {Bailin},
  {Dedes}, {Janowiecki}, {Gibson}, {Murphy}, {Nakanishi}, \&
  {Newton-McGee}}]{mcclure09}
{McClure-Griffiths}, N.~M., {Pisano}, D.~J., {Calabretta}, M.~R., {et~al.}
  2009, \apjs, 181, 398

\bibitem[{{Meyer} {et~al.}(2004){Meyer}, {Zwaan}, {Webster}, {Staveley-Smith},
  {Ryan-Weber}, {Drinkwater}, {Barnes}, {Howlett}, {Kilborn}, {Stevens},
  {Waugh}, {Pierce}, {Bhathal}, {de Blok}, {Disney}, {Ekers}, {Freeman},
  {Garcia}, {Gibson}, {Harnett}, {Henning}, {Jerjen}, {Kesteven}, {Knezek},
  {Koribalski}, {Mader}, {Marquarding}, {Minchin}, {O'Brien}, {Oosterloo},
  {Price}, {Putman}, {Ryder}, {Sadler}, {Stewart}, {Stootman}, \&
  {Wright}}]{meyer04}
{Meyer}, M.~J., {Zwaan}, M.~A., {Webster}, R.~L., {et~al.} 2004, \mnras, 350,
  1195

\bibitem[{{Peek} \& {Heiles}(2008)}]{peek08b}
{Peek}, J.~E.~G. \& {Heiles}, C. 2008, arXiv e-prints, 0810.1283

\bibitem[{{Stanko} {et~al.}(2005){Stanko}, {Klein}, \& {Kerp}}]{stanko05}
{Stanko}, S., {Klein}, B., \& {Kerp}, J. 2005, \aap, 436, 391

\bibitem[{{Taylor} {et~al.}(2003){Taylor}, {Gibson}, {Peracaula}, {Martin},
  {Landecker}, {Brunt}, {Dewdney}, {Dougherty}, {Gray}, {Higgs}, {Kerton},
  {Knee}, {Kothes}, {Purton}, {Uyaniker}, {Wallace}, {Willis}, \&
  {Durand}}]{taylor03}
{Taylor}, A.~R., {Gibson}, S.~J., {Peracaula}, M., {et~al.} 2003, \aj, 125,
  3145

\bibitem[{{Winkel} \& {Kerp}(2007)}]{winkel07b}
{Winkel}, B. \& {Kerp}, J. 2007, \apjs, 173, 166

\bibitem[{{Winkel} {et~al.}(2007){Winkel}, {Kerp}, \& {Stanko}}]{winkel07a}
{Winkel}, B., {Kerp}, J., \& {Stanko}, S. 2007, Astronomical Notes, 328, 68

\end{thebibliography}




\clearpage

\end{document}